\documentclass{elsarticle}
\usepackage{amsmath,amssymb,amsthm,graphicx,enumerate,url}
\usepackage[small,bf]{caption}
\usepackage{color}
\usepackage{subfig}
\usepackage{multirow}
\usepackage{dcolumn}
\usepackage{tabularx}
\usepackage{threeparttable}
\usepackage{booktabs}
\usepackage{multirow}
\usepackage{array}
\biboptions{sort&compress}
\usepackage[letterpaper]{geometry}
\newcount\Comments  
\usepackage{color}
\definecolor{darkgreen}{rgb}{0,0.5,0}
\definecolor{purple}{rgb}{1,0,1}
\newcommand{\kibitz}[2]{\ifnum\Comments=0\textcolor{#1}{#2}\fi}

\theoremstyle{plain}

\theoremstyle{definition}

\theoremstyle{remark}
\newtheorem{rem}{Remark}

\newcommand{\pd}[2]{\frac{\partial#1}{\partial#2}}
\newcommand{\ud}[1]{\,\mathrm{d}#1}

\graphicspath{ {./images/} }

\usepackage{tabularx}

\title{A Collapsed Generalized Aw-Rascle-Zhang Model and Its Model Accuracy}

\author{Shimao Fan\fnref{1}}
\fntext[1]{S. Fan is a Postdoctoral Researcher in the Department of Civil and Environmental Engineering, University of Illinois at Urbana Champaign,  Urbana, IL 61801, USA.}

\author{Ye Sun\fnref{2}}
\fntext[2]{Y. Sun is a PhD student in the Department of Civil and Environmental Engineering \& Coordinated Science Laboratory, University of Illinois at Urbana Champaign,  Urbana, IL 61801, USA.}

\author{Benedetto Piccoli\fnref{3}}
\fntext[3]{B. Piccoli is a Professor in the Department of Mathematical Sciences, Rutgers University, Camden, Camden, NJ 08102, USA.}

\author{Benjamin Seibold\fnref{4}}
\fntext[4]{B. Seibold is an Associate Professor in the Department of Mathematics, Temple University, Philadelphia PA, 19122, USA.}

\author{Daniel B. Work\corref{mycorrespondingauthor}\fnref{5}}
\fntext[5]{D. Work is an Assistant Professor in the Department of Civil and Environmental Engineering \& Coordinated Science Laboratory, University of Illinois at Urbana Champaign,  Urbana, IL 61801, USA. Email: dbwork@illinois.edu}
\cortext[mycorrespondingauthor]{Corresponding author}
\cortext[mycorrespondingauthor]{This material is based upon work supported by the National Science Foundation under Grant Nos. CNS-1446715 (B.P.), CNS-1446690 (B.S.). \& CMMI-1351717 (D.W.).}

\date{}
\usepackage{lipsum}
\makeatletter
\def\ps@pprintTitle{%
 \let\@oddhead\@empty
 \let\@evenhead\@empty
 \def\@oddfoot{}%
 \let\@evenfoot\@oddfoot}
\makeatother
\begin{document} 
\begin{frontmatter}

\begin{abstract}

This work presents a collapsed generalized Aw-Rascle-Zhang (CGARZ) model, which fits into a generic second order model (GSOM) framework. GSOMs augment the evolution of the traffic density by a second state variable characterizing a property of vehicles or drivers. A cell transmission model for the numerical solution of GSOMs is derived, which is based on analyzing the sending and receiving functions of the traffic density and total property. The predictive accuracy of the CGARZ model is then compared to the classical first-order LWR and four second-order models. To that end, a systematic approach to calibrate model parameters from sensor flow-density data is introduced and applied to all models studied. The comparative model validation is conducted using two types of field data: vehicle trajectory data, and loop detector data. It is shown that the CGARZ model provides an intriguing combination of simple model dynamics in the free-flow regime and a representation of the spread of flow--density data in the congested regime, while possessing a competitive prediction accuracy.
\end{abstract}
\end{frontmatter}

\section{Introduction}
\label{sec:intro}
Macroscopic traffic flow models play a critical role in modern traffic control and estimation techniques \cite{BlandinCoqueBayen2012,KotsialosPapageorgiou2001}. These models include the \textit{Lighthill-Whitham-Richards} (LWR) \cite{LighthillWhitham1955,Richards1956} model, \textit{phase transition} (PT) models \cite{blandin2013phase,BlandinWorkGoatinPiccoliBayen2011,Colombo2003,colombo2007global},  the \textit{Aw-Rascle-Zhang} (ARZ) model \cite{AwRascle2000,BerthelinDegondDelitalaRascle2008,Greenberg2001,Zhang2002}, the \textit{generalized ARZ} (GARZ) model \cite{fan2014comparative}, and the \textit{collapsed GARZ} (CGARZ) model \cite{Fan2013} introduced in this work. The LWR, ARZ, GARZ, and CGARZ models fit into the \textit{generic second order model} (GSOM) \cite{LebacqueMammarHajSalem2007} framework:
\begin{equation}
\label{eq:generic}
\begin{split}
\rho_t + (\rho v)_x &= 0 ,\\
w_t + v w_x & = 0 ,\\
\text{with}\quad v &= V(\rho,w) .
\end{split}
\end{equation}
Here $\rho(x,t)$, $v(x,t)$, and $w(x,t)$ represent the traffic density, velocity, and \textit{property}, respectively, that depend on both space (position along the road) $x$ and time $t$. The property $w$ can have various meanings, for instance the fraction of special vehicles in the total traffic stream (e.g., trucks or autonomous vehicles) \cite{fan2015heterogeneous}, ``aggressivity"  \cite{fan2014comparative}, ``desired spacing" \cite{zhang2009conserved}, or ``perturbation from equilibrium" \cite{BlandinWorkGoatinPiccoliBayen2011}. The first equation of \eqref{eq:generic} describes the conservation of vehicles, while the second equation of \eqref{eq:generic} indicates that the property $w$ is advected with the vehicles at speed $v$. Consequently, when $w$ is interpreted as a vehicle class fraction, it implies no systematic overtaking between the vehicle classes.
The quantity $w$ is used to relate driver properties to the flow--density curves. Thus, the GSOM possesses a family of \textit{fundamental diagrams} (FDs), $Q(\rho,w) = \rho V(\rho,w)$, parametrized by $w$. 

The model \eqref{eq:generic} is given in advection form. To admit a discretization that guarantees the conservation of the \textit{total property} $y=\rho w$ (see Section~\ref{section:numerical}), the model can be written in conservative form:
\begin{equation}
\label{eq:generic_conserv}
\begin{split}
\rho_t + (\rho v)_x &= 0 ,\\
y_t + (y v)_x & = 0 ,\\
\text{with}\quad y = \rho w ,\quad v &= V(\rho,y/\rho) .
\end{split}
\end{equation}
Traffic models of the form \eqref{eq:generic_conserv} were originally motivated in analogy to gas dynamics \cite{AwRascle2000,Colombo2003}. Based on the fact that $w$ is advected with the vehicle flow and $\rho$ is conserved, it follows that the quantity $y = \rho w$ is also conserved. 


When all drivers have the same property ($w = \bar{w} = \text{const.}$), the GSOM reduces to the LWR model \cite{LighthillWhitham1955,Richards1956}
\begin{equation}\label{eq:lwr}
\rho_t + \left(\rho V(\rho)\right)_x = 0,
\end{equation}
where the velocity $V(\rho) = V(\rho,\bar{w})$ now depends on the density only. Hence, the LWR model is included as a special case of the GSOM class \cite{LebacqueMammarHajSalem2007,FanSeibold2013}. In \eqref{eq:lwr}, the unique flow--density relationship $Q(\rho) = \rho V(\rho)$ defines the fundamental diagram.

%
In the ARZ \cite{AwRascle2000, BerthelinDegondDelitalaRascle2008,Greenberg2001,Zhang2002} and the GARZ models \cite{fan2014comparative}, the traffic velocity and flow vary with property $w$ for all $\rho\in [0,\rho_\text{max}]$, where $\rho_\text{max}$ is the maximum traffic density. In other words, each property value $w$ corresponds to its own distinct FD curve $v_w(\rho) = V(\rho,w)$. For instance, in the ARZ model \cite{Zhang2002}, the velocity function is defined as
\begin{equation}
\label{eq:velocity_arz}
V(\rho,w) = V_\text{eq}(\rho)+\left(w-V_\text{eq}(0)\right),\quad \text{for $\rho\in [0,\rho_\text{max}]$,}
\end{equation}
where $V_\text{eq}(\rho)$ represents the \emph{equilibrium velocity function}, and the associated flow--density function $Q_\text{eq}(\rho) = \rho V_\text{eq}(\rho)$ is an \emph{equilibrium FD}. According to \eqref{eq:velocity_arz}, a family of velocity curves for the ARZ model is generated by shifting the equilibrium velocity curve vertically with $V(0,w) = w$. Thus $\pd{V(\rho,w)}{w} = 1>0$, which means that the traffic velocity always varies with $w$.

However, fundamental diagram measurements of real traffic exhibit a distinct structure: in the congested regime, the flow rates corresponding to a certain density possess a significant spread; in contrast, the spread of the data in the free-flow regime is substantially smaller. In fact, various studies of traffic models \cite{Kerner2000a,Kerner2000b} indicate that the origin of the spreads in the two regimes is of fundamentally different nature: in the congested regime, spread results largely from systematic interactions between vehicles, and from non-equilibrium effects such as unstable, or bi-stable, traffic flow phenomena \cite{BandoHesebeNakayama1995,FlynnKasimovNaveRosalesSeibold2009,SeiboldFlynnKasimovRosales2013}.
In contrast, in the free-flow regime, interactions between vehicles are weak, and the spread in real data describing aggregate traffic quantities is driven by measurement errors and the small number of vehicles over which the aggregate quantiles are computed. The premise of macroscopic traffic flow models is that they do not describe (in fact: average out) measurement noise and the behavior of an individual driver, but that they \emph{do} describe the aggregate dynamics based on the collective vehicle interactions.

Motivated by this premise, attempts have been made to develop traffic models that possess a single-valued fundamental diagram for low densities, and a multi-valued fundamental diagram in the congested regime. One of the most prominent ideas is to use \textit{phase transition} (PT) models \cite{colombo2007global, BlandinWorkGoatinPiccoliBayen2011,blandin2013phase,Colombo2003}, which describe free-flow traffic (i.e., low density traffic with small speed variations for a fixed density) via first order dynamics (i.e., the LWR model) and congestion via second order dynamics.
While phase transition models do capture the spread of observed data in congestion, the explicit introduction of a phase transition increases the mathematical complexity of the model, and consequently it has not yet seen widespread use in engineering applications such as estimation and control.

In this article we present and validate an alternative way to construct a macroscopic model (i.e., the CGARZ) that possesses a single-valued fundamental diagram in free-flow, and a multi-valued regime in congestion. In the CGARZ model, it is assumed that the property $w$ does not influence the actual traffic behavior in the low density regime. This means that vehicles may possess different properties, but the velocity and flow in free-flow is not affected by $w$, i.e., $\pd{V(\rho,w)}{w} = 0$. The CGARZ model is a special form of the GARZ model that collapses the flow--density curves into a single curve in the free-flow region (see Figure~\ref{fig:cgarz_flux} for the CGARZ flow--density diagram). 
Consequently, the analytical results of the ARZ and the GARZ models \cite{AwRascle2000,fan2014comparative,LebacqueMammarHajSalem2007} transfer over to the CGARZ model.

One may note the important distinction between phase transition models and the CGARZ model. In the PT models, a scalar conservation law is used to model traffic in the free-flow phase, denoted $\Omega_\text{f}$, and a GSOM is employed in the congested phase $\Omega_\text{c}$. Thus, the property $w$ is conserved in congestion, but it is undefined in free-flow. Specifically, the information carried by $w$ gets lost when traffic flow transitions into free-flow. Depending on the interpretation/meaning of the property (e.g., fraction of trucks in the vehicle stream), this can be a distinct disadvantage of the model.

Hence, the CGARZ model combines desirable features of both classes of models (PT and GSOM): \textit{i}) it is appropriate in modeling distinct behaviors in free-flow vs.~congested phases, which agrees with Kerner's empirical observation \cite{Kerner2000a,Kerner2000b}; \textit{ii}) it conserves the total property $y=\rho w$ in free-flow and in congestion; and \textit{iii}) it avoids the complicated analytical work required in a PT model. The main properties of the CGARZ model are presented in Section~\ref{section:cgarz}.


Then, in Section~\ref{section:numerical}, a numerical solver for the CGARZ model is instantiated, which is a generalization of the \emph{cell transmission model} (CTM)  \cite{daganzo1994cell} for second order traffic flow models, called here the 2CTM. The CTM is a Godunov discretization~\cite{Godunov1959,LeVeque2002} of the LWR model that explains the discrete evolution of traffic state by defining a physical interpretation of how traffic flows from one cell to the next. Specifically, in the presented solver we define the sending and receiving (equivalently the demand and supply) functions in terms of the upstream and downstream traffic quantities.

Finally, in Section~\ref{section:validation}, a systematic approach to construct the CGARZ model from sensor data is introduced, which is also applicable to all the other models mentioned above. 
Then, then CGARZ model is validated, by following the test framework introduced in \cite{FanSeibold2013,fan2014comparative}, and compared with existing macroscopic traffic models, such as a PT model and other models that fit in to the generic framework of GSOM.

\section{The Collapsed Generalized ARZ Model}\label{section:cgarz}

\subsection{Outline of the CGARZ Model}

\begin{figure}[t]
\begin{center}
  \subfloat[]{\label{fig:cgarz_flux}\includegraphics[width=0.33\textwidth]{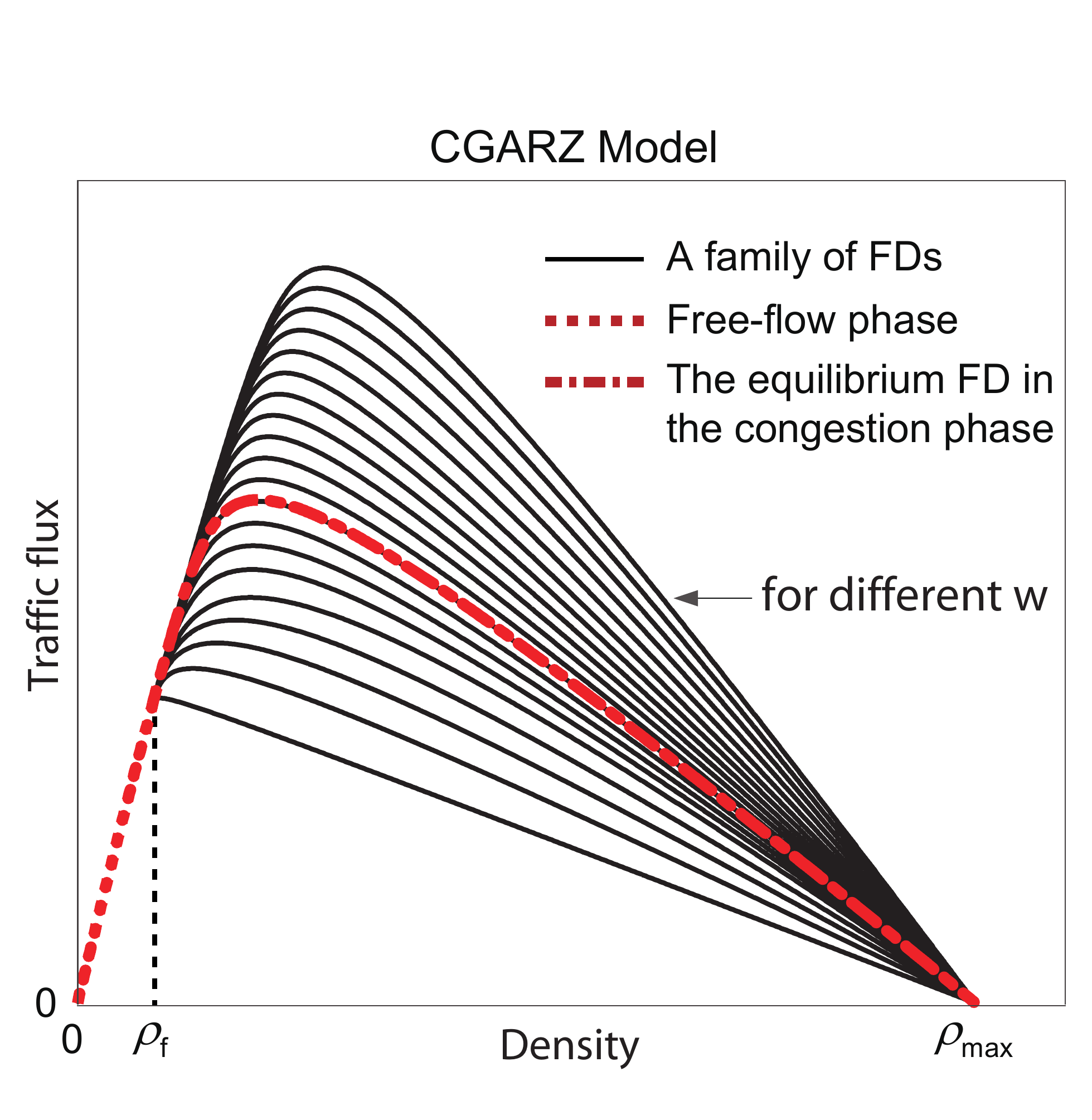}}
~\qquad
  \subfloat[]{\label{fig:cgarz_velocity}\includegraphics[width=0.33\textwidth]{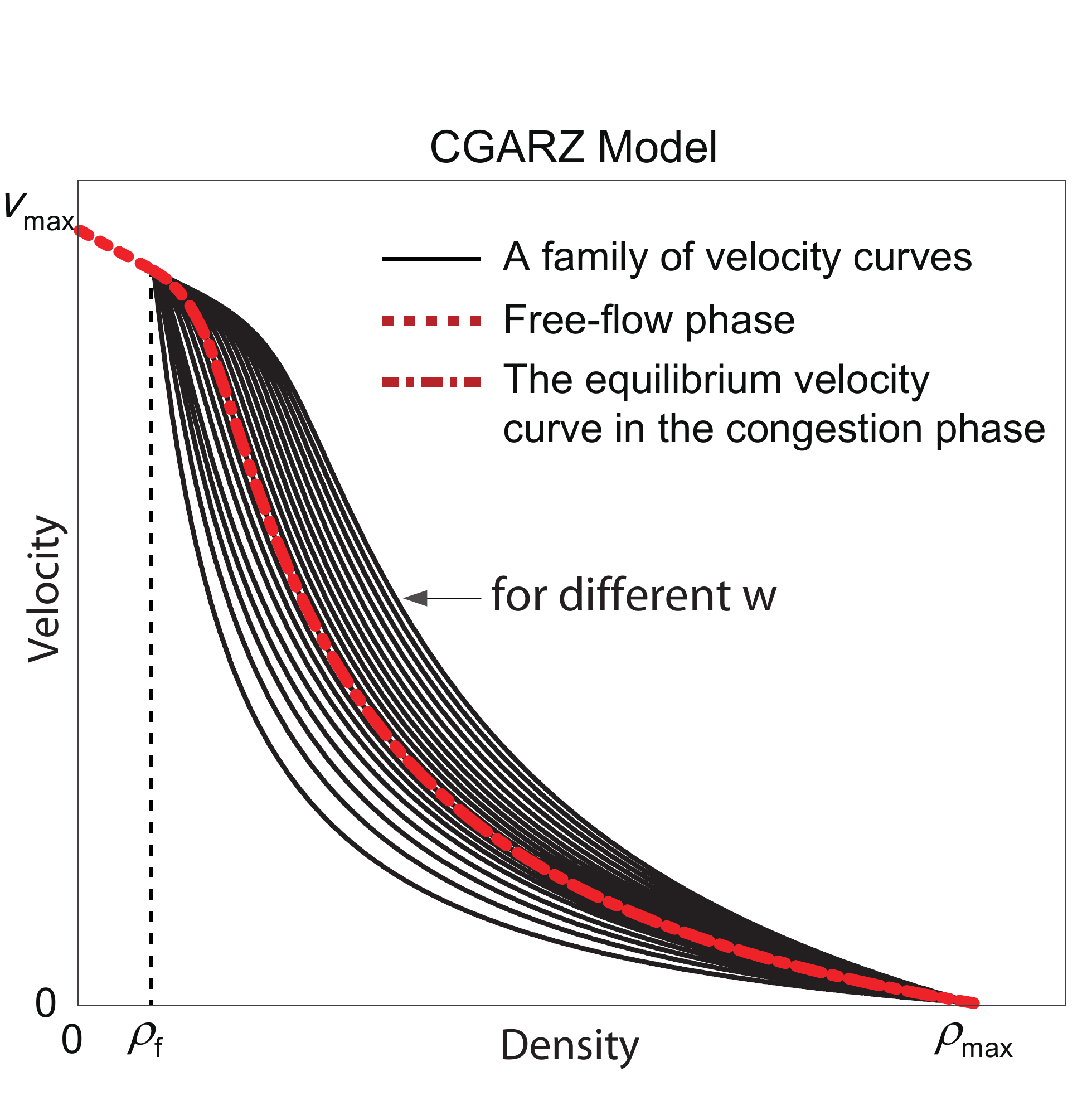}}
\caption{(a) Flow--density curves of the CGARZ model. Although the free-flow flow--density relationship appears to be linear, it is in fact curved. (b) Velocity--density curves of the CGARZ model. In the model, the velocity is strictly decreasing with respect to the density, and $v_{\text{max}}$ is the maximum traffic velocity. The equilibrium fundamental diagram and the equilibrium velocity function are shown in red.}
\label{fig:cgarz}
\end{center}
\end{figure}

In this section, the main properties of the CCARZ model are presented. A more detailed description of the model and its properties is available in~\cite{Fan2013}. To capture distinct behaviors of traffic flow in different regimes, the flow--density curves of the GARZ model are collapsed to a single curve for $0\le\rho\le\rho_{\text{f}}$, where $\rho_{\text{f}}$ is a \textit{free-flow threshold density} (see Figure~\ref{fig:cgarz}), resulting in the CGARZ model that fits the generic second order model (GSOM) framework \eqref{eq:generic_conserv}. The flux function of the CGARZ model (see Figure~\ref{fig:cgarz_flux}) is given as:
\begin{equation}
\label{fd_cgarz}
 Q(\rho,w) = 	\left\{\begin{array}{rl}
	        Q_{\text{f}}(\rho) , &\quad\text{ if}~ 0\le\rho\le\rho_{\text{f}} ,\\
	         Q_{\text{c}}(\rho,w) , &\quad\text{ if}~ \rho_{\text{f}}<\rho\le \rho_{\text{max}} ,\end{array} \right.
\end{equation}
where $\rho_{\text{max}}$ is the maximum traffic density (i.e., the density of traffic when vehicles are at rest with minimal spacing between each vehicle), $Q_{\text{f}}(\rho)$ is the flux function in free-flow, and $Q_\text{c}(\rho,w)$ is a family of flow--density curves in congestion (see Section~\ref{sec:models_FD} for an example of $Q_{\text{c}}(\rho,w)$ used in the data fitting analysis presented thereafter). It should be stressed that the free-flow threshold density $\rho_{\text{f}}$ is defined as the onset of spread in the FD, and not as the density at which maximum flow is achieved.

In this work, the Greenshields model is applied in free-flow: 
\begin{equation}
\label{eq:freediagram}
Q_{\text{f}}(\rho) = v_\text{max}\,\rho\left(1-\frac{\rho}{\tilde{\rho}_\text{max}}\right) ,
\end{equation}
where $v_\text{max}$ is the maximum velocity, and $\tilde{\rho}_\text{max}$ is a shape parameter to control the curvature of the fundamental diagram in free-flow. Note that $\tilde{\rho}_\text{max}$ is not the maximum density since \eqref{eq:freediagram} is only valid for $0\le\rho\le\rho_{\text{f}}$. 

The flux function $Q(\rho,w)$ is designed with the following properties:
\begin{enumerate}
 \item Flow--density curves have a common $\rho_{\text{max}}$ independent of $w$, i.e., $Q(\rho_\text{max},w)=0$ for all $w$.
 \item The flux function is strictly concave with respect to $\rho$, i.e., $\frac{\partial^2 Q(\rho,w)}{\partial \rho^2}  <0$, for $\rho\in[0,\rho_\text{max})$.
 \item The flux function is continuously differentiable in $\rho$, i.e., $Q(\rho,w)\in \mathcal{C}^1$, for each $w$.
 \item If a flow--density curve has a larger property value $w_1>w_2$, then $Q(\rho,w_1)$ lies completely above $Q(\rho,w_2)$ for $\rho_{\text{f}}<\rho < \rho_{\text{max}}$, i.e., $\frac{\partial Q(\rho,w)}{\partial w}>0$, if $\rho_{\text{f}}<\rho < \rho_{\text{max}}$. 
\end{enumerate}

The flux function \eqref{fd_cgarz} induces a velocity function
\begin{equation}\label{eq:vel_definition}
 V(\rho,w) = 	\left\{\begin{array}{rl}
	        V_{\text{f}}(\rho) , &\quad\text{ if}~ 0\le\rho\le\rho_{\text{f}} ,\\
	         V_{\text{c}}(\rho,w) , &\quad\text{ if}~ \rho_{\text{f}}<\rho\le \rho_{\text{max}} .\end{array} \right.
\end{equation}
where $V_{\text{f}}(\rho) = v_\text{max}\left(1-\frac{\rho}{\tilde{\rho}_\text{max}}\right)$ and $V_{\text{c}}(\rho,w) = \frac{Q_\text{c}(\rho,w)}{\rho}$ (see Figure~\ref{fig:cgarz_velocity}).
Based on the properties of the flux function, the velocity function \eqref{eq:vel_definition} is in $\mathcal{C}^1$ and is strictly decreasing in density. Thus, $v_{\text{max}}$ is the unique maximum velocity independent of $w$.
The features of the velocity function are summarized as follows:
\begin{enumerate}
 \item Vehicles never go backwards, i.e., $V(\rho,w)\ge 0$.
 \item Vehicles possess a unique density at which the velocity is zero, i.e., $V(\rho_{\text{max}},w) = 0$.
 \item In the free-flow regime, the traffic velocity is independent of the property $w$, i.e., $\pd{V(\rho,w)}{w} = 0$, if $0\le\rho\le\rho_{\text{f}}$.
 \item The traffic velocity is increasing with respect to the property in the congested region, i.e., $\frac{\partial V(\rho,w)}{\partial w}>0$, if $\rho_{\text{f}}<\rho\le \rho_{\text{max}}$.
\end{enumerate}

In summary, the CGARZ model in conservation form reads as
\begin{equation}
\label{eq:cgarz}
\left\{\begin{array}{rl}
\rho_t + (\rho v)_x &= 0 ,\\
y_t + (y v)_x & = 0 ,\\
v=V(\rho,y/\rho)&=\left\{\begin{array}{rl}
	        V_{\text{f}}(\rho) , &~\text{ if}~ 0\le\rho\le\rho_{\text{f}} ,\\
	         V_{\text{c}}(\rho,y/\rho) , &~\text{ if}~ \rho_{\text{f}}<\rho\le \rho_{\text{max}} .\end{array} \right.
\end{array} \right.	
\end{equation}
%
%
Note that due to the collapsing, the dynamics of the traffic flow in the free-flow region $0\le\rho\le\rho_{\text{f}}$ are effectively determined by the first-order LWR model. However, the key contrast to PT models is that in the CGARZ model, the property field $w$ is still being tracked while in free-flow, even though the actual traffic dynamics are not affected by it.
%




\subsection{Inverse Functions in the CGARZ Model}\label{section:inverse}
The model dynamics are completely determined by the velocity function \eqref{eq:vel_definition}, which, given $\rho$ and $w$, determines the velocity $v$. However, its numerical approximation (Section~\ref{section:numerical}), and applying the model in a validation framework (Section~\ref{section:validation}) also require the two corresponding inverse functions:
\begin{enumerate}
\item Given $v$ and $w$, determine $\rho = G(v,w)$, s.t.~$v=V(\rho,w)$. This function is needed to identify an intermediate traffic density in the 2CTM (see Section~\ref{section:numerical}). We assume that there is a lower and upper bound for $w$, i.e., $0<w_\text{min}\le w\le w_\text{max}$. Because $\pd{V(\rho,w)}{\rho} < 0$, the function $V(\rho,w)$ uniquely defines the function
\begin{equation}
\label{eq:inverse_function_G}
G:\mathcal{D}_G \longrightarrow [0,\rho_\text{max})
\text{~~where~~}
\mathcal{D}_G = \{(v,w) \,|\,  0<v\le v_{\text{max}},  w_\text{min}\le w\le w_\text{max}\}.
\end{equation}
\item Given $\rho$ and $v$, determine $w = W(\rho,v)$, s.t.~$v = V(\rho,w)$. This function is needed to construct the property $w$ (which is not directly measured, but needed to specify boundary conditions) from the measurement data $\rho$ and $v$. Here the collapsing induces a difficulty: in contrast to the (non-collapsed) GARZ model \cite{fan2014comparative}, the CGARZ model does not possess uniquely defined inverse function (because of the collapsing). As a consequence, additional modeling choices are needed to define the function $W$. We do so as follows. For congested densities, the inverse mapping
\begin{equation*}
W_\text{c}:\mathcal{D}_{W,\text{c}} \longrightarrow [w_{\text{min}},w_\text{max}]
\text{~~where~~}
\mathcal{D}_{W,\text{c}} = \{(\rho,v) \,|\,  \rho_\text{f}<\rho\le \rho_\text{max},
V(\rho,w_\text{min})\le v\le V(\rho,w_\text{max})\}
\end{equation*}
is uniquely determined through $V(\rho,w)$. Moreover, in free-flow, we define the mapping
\begin{equation*}
W_\text{f}:\mathcal{D}_{W,\text{f}} \longrightarrow [w_{\text{min}},w_\text{max}]
\text{~~where~~}
\mathcal{D}_{W,\text{f}} = \{(\rho,v) \,|\,  0\le \rho \le \rho_\text{f},
0\le v\le v_\text{max}\}
\end{equation*}
by simply assigning $W_\text{f}(\rho,v) = w_{\text{eq}}$, where $w_{\text{eq}} \in (w_\text{min},w_\text{max})$ corresponds to the equilibrium curve. The complete function then reads as
\begin{equation*}
W(\rho,v) = \begin{cases}
W_\text{f}(\rho,v) & 0\le \rho \le \rho_\text{f} \\
W_\text{c}(\rho,v) & \rho_\text{f}<\rho\le\rho_\text{max}\;.
\end{cases}
\end{equation*}
\end{enumerate}

\section{Numerical Method}\label{section:numerical}
In this section, a numerical solver, the 2CTM, is presented in analogy to the classical CTM \cite{daganzo1994cell}. The 2CTM is applicable to all the models that fit into the GSOM framework \eqref{eq:generic_conserv}, in particular the CGARZ model. The classical CTM is mathematically equivalent to a Godunov approximation \cite{Godunov1959} of the LWR model. However, its formulation in terms of sending (demand) and receiving (supply) functions provides important insight for practical implementations, for network coupling conditions \cite{GaravelloPiccoli2006,GaravelloPiccoli2016}, and for control \cite{HertyKlar2003}.

In line with \cite{lebacque2005second}, we formulate the new 2CTM based on a Godunov approximation of the CGARZ/GSOM, with a central focus on sending and receiving functions. In particular, a concrete physical interpretation of the \emph{intermediate state}, that arises in the Riemann problem \cite{AwRascle2000,Zhang2002,Colombo2003,lebacque2007aw} due to the second-order model dynamics, is provided. The resulting formulation yields a discrete model in which the flux across cell interfaces is based on the sending capacity of the upstream cell and the receiving capacity at the downstream cell. Before presenting the 2CTM, we first summarize the key properties of the classical CTM.



\subsection{Cell Transmission Model}
\begin{figure}[t]
\centering
  \includegraphics[width=0.75\textwidth]{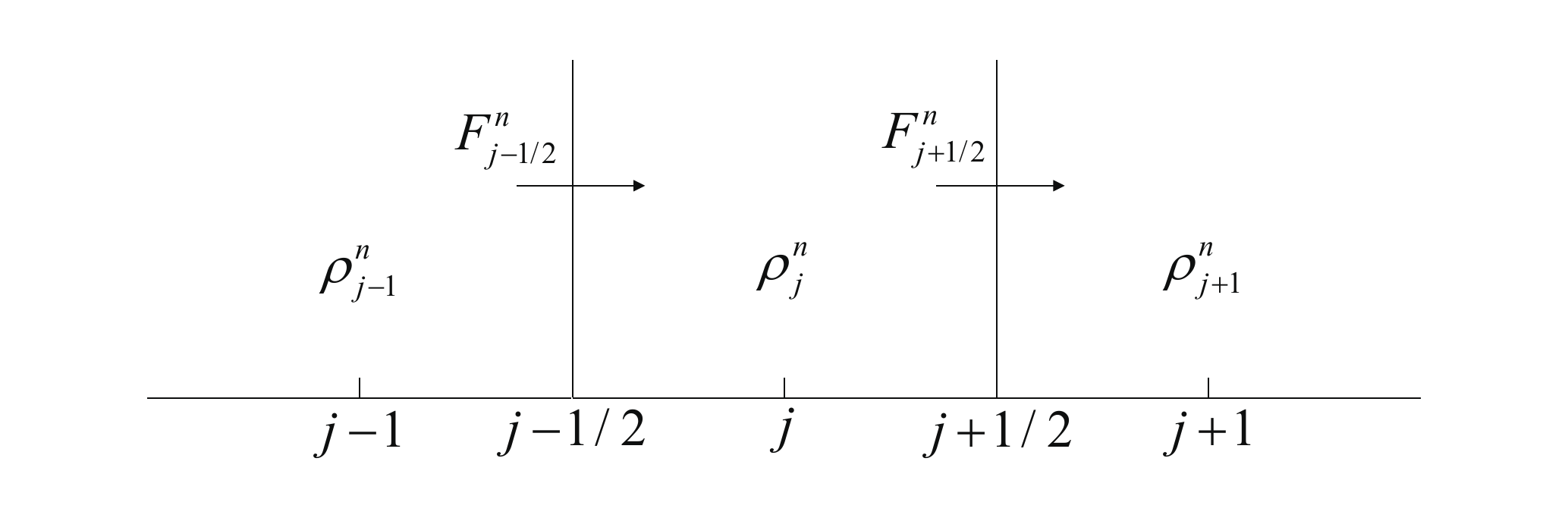}
\caption{The cell transmission model.}
\label{fig:ctm}
\end{figure}

The CTM \cite{daganzo1994cell} is based on the integral form of the LWR model \eqref{eq:lwr}
\begin{equation*}
\label{eq:integral}
 \frac{d}{d t}\int_{a}^{b}{\rho(x,t)}dx = F(a,t) - F(b,t) ,
\end{equation*}
where $F(a,t)$ and $F(b,t)$ represent the incoming and outgoing fluxes at the boundaries of a cell, respectively, and the integral on the left hand side is the number of vehicles on the road segment $x\in[a,b]$.

The CTM discretizes the space into cells with size $\Delta x$, and studies each cell by examining its inflow and outflow over the time interval $\Delta t$. We consider three adjacent cells with initial densities $\rho_{j-1}^{n}$, $\rho_{j}^{n}$, and $\rho_{j+1}^{n}$ at the time $t = n\Delta t$, and study the evolution of traffic density in the $j$-th cell (see Figure~\ref{fig:ctm}). The key features of the CTM are listed as follows:
\begin{enumerate}
 \item The evolution equation is given by the Godunov method \cite{Godunov1959}:
\begin{equation}
\label{eq:sd_1}
 \rho^{n+1}_j = \rho^{n}_j + \frac{\Delta t}{\Delta x} \left(F^{n}_{j-1/2} - F^{n}_{j+1/2}\right) ,
\end{equation}
where $F^{n}_{j-1/2}$ and $F^{n}_{j+1/2}$ are the inflow and outflow of the $j$-th cell over the time interval $t\in[n\Delta t, (n+1)\Delta t)$.

\item $F^{n}_{j-1/2}$ and $F^{n}_{j+1/2}$ are determined by the minimum of the vehicles available to be sent from the upstream, and the availability of the downstream cell to receive vehicles:
\begin{equation*}
 \label{eq:flow_1}
    F^{n}_{j-1/2} = \min\left\{ S(\rho_{j-1}^{n}) ,  R(\rho_{j}^{n})\right\} ,\qquad F^{n}_{j+1/2} = \min\left\{ S(\rho_{j}^{n}) ,  R(\rho_{j+1}^{n})\right\} ,
\end{equation*}
where $S(\cdot)$ and $R(\cdot)$ are the so-called \emph{sending} and \emph{receiving} functions.

\item The sending and receiving functions are defined through the fundamental diagram $Q(\rho)$
\begin{equation}
 \label{eq:ctm}
   S(\rho) = \left\{\begin{array}{rl} Q(\rho) , &\quad\text{if } \rho\le \rho_{\text{c}} , \\
	          Q^{\text{max}} , &\quad\text{if }  \rho > \rho_{\text{c}} ,\end{array} \right.\qquad
   R(\rho) = \left\{\begin{array}{rl} Q^{\text{max}} , &\quad\text{if }  \rho \le \rho_{\text{c}} , \\
	          Q(\rho) , &\quad\text{if } \rho> \rho_{\text{c}} ,\end{array} \right. \\		
\end{equation}
where $\rho_{\text{c}}$ denotes the critical density where $Q^{\text{max}}$ is obtained.
\end{enumerate}

\subsection{Second Order Cell Transmission Model}
\begin{figure}[t]
\centering
  \includegraphics[width=0.75\textwidth]{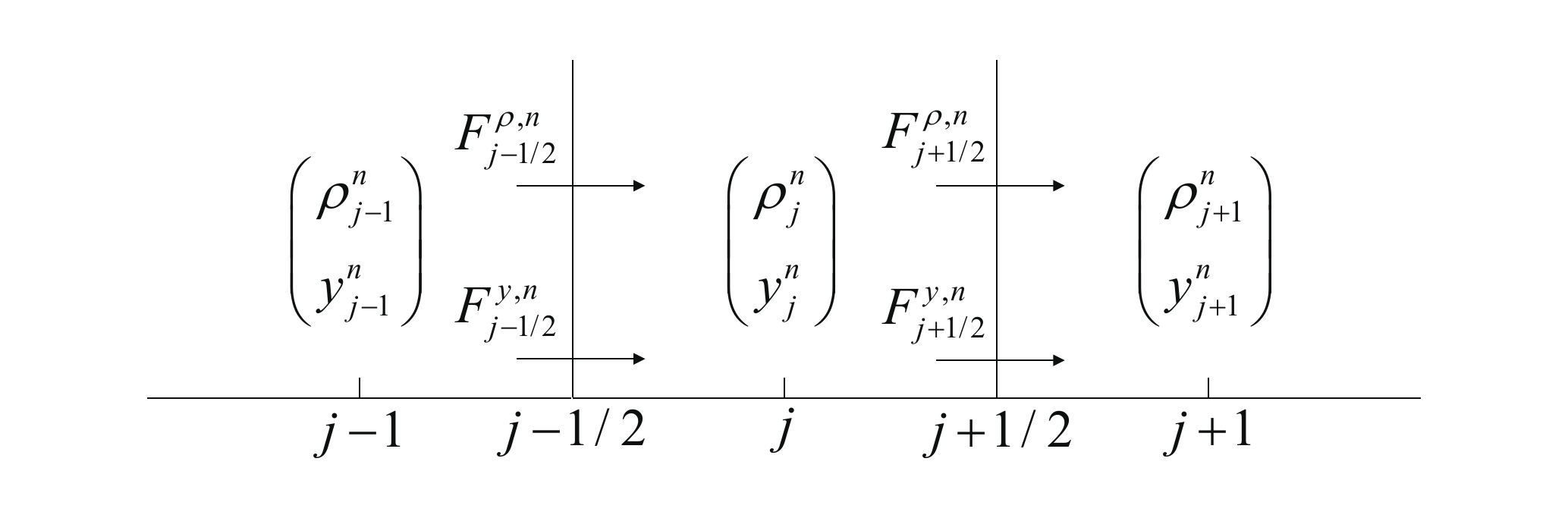}
\caption{Second order cell transmission model.}
\label{fig:2ndctm}
\end{figure}

Since the CGARZ model together with the LWR, ARZ and GARZ models fit into the GSOM framework, the 2CTM is designed based on the GSOM \eqref{eq:generic_conserv}. For a system of conservation laws \eqref{eq:generic_conserv}, the initial traffic states in three adjacent cells are vectors $u^{n}_{j-1}=\left(\rho_{j-1}^{n},\rho_{j-1}^{n}w^{n}_{j-1}\right)$, $u^{n}_{j}=\left(\rho_{j}^{n},\rho_{j}^{n}w^{n}_{j}\right)$, and $u^{n}_{j+1}=\left(\rho_{j+1}^{n},\rho_{j+1}^{n}w^{n}_{j+1}\right)$  at the time $t = n\Delta t$ (see Figure~\ref{fig:2ndctm}). By applying the Godunov scheme \cite{Godunov1959,LeVeque2002} to \eqref{eq:generic_conserv}, the 2CTM has the following form:
\begin{equation}
\label{eq:sd_original}
 \begin{split}
 \rho^{n+1}_j &= \rho^{n}_j + \frac{\Delta t}{\Delta x} \left(F^{\rho,n}_{j-1/2} - F^{\rho,n}_{j+1/2}\right) ,\\
  y^{n+1}_j &= y^{n}_j + \frac{\Delta t}{\Delta x} \left(F^{y,n}_{j-1/2} - F^{y,n}_{j+1/2}\right) ,
 \end{split}
\end{equation}
which provides evolution equations for both conserved quantities, traffic density $\rho$ and total property $y = \rho w$. Here, $F_{j-1/2}^{\rho,n}$ (resp.~$F_{j+1/2}^{\rho,n}$) and $F_{j-1/2}^{y,n}$ (resp.~$F_{j+1/2}^{y,n}$) are the inflows (resp.~outflows) of $\rho$ and $y$ over the time interval $t\in[n\Delta t, (n+1)\Delta t)$, respectively.

To determine $F^{\rho}$ and $F^{y}$, it is important to note that the two flows are related. Since the property $w$ is always advected with vehicle flow $F^{\rho}$, the flow of total property $F^{y}$ is computed by multiplying the property $w$ from the upstream to the flow of vehicles:
\begin{equation*}
 \label{eq:flow_y}
     F^{y,n}_{j-1/2} = w^{n}_{j-1} F^{\rho,n}_{j-1/2} ,\quad \text{and}\quad F^{y,n}_{j+1/2} = w^{n}_{j} F^{\rho,n}_{j+1/2} ,
\end{equation*}
where $w^{n}_{j-1}$ and $w^{n}_{j}$ are the properties of vehicles at the $(j-1)$-st and the $j$-th cells. Thus, the update equations of the 2CTM \eqref{eq:sd_original} simplify to
\begin{equation}
\label{eq:sd_2}
 \begin{split}
 \rho^{n+1}_j &= \rho^{n}_j + \frac{\Delta t}{\Delta x} \left(F^{\rho,n}_{j-1/2} - F^{\rho,n}_{j+1/2}\right) ,\\
 y^{n+1}_j &= y^{n}_j + \frac{\Delta t}{\Delta x} \left(w^n_{j-1} F^{\rho,n}_{j-1/2} -w^n_{j} F^{\rho,n}_{j+1/2}\right) .
 \end{split}
\end{equation}

Next, the vehicle flow through a cell boundary is the minimum of the sending and receiving functions like the CTM. To complete the scheme \eqref{eq:sd_2}, it is sufficient to define the sending and receiving functions for $F^{\rho}$ similarly to the CTM.

\begin{rem}
In the special case that all vehicles have the same property, i.e., $w(x,t) = \bar{w}$, the CGARZ model reduces to the LWR model. Likewise, the update equation for $y$ becomes identical to that for $\rho$,
 \begin{equation*}
  \bar{w} \rho^{n+1}_j = \bar{w}\rho^{n}_j + \frac{\Delta t}{\Delta x} \left(\bar{w}F^{\rho,n}_{j-1/2} - \bar{w}F^{\rho,n}_{j+1/2}\right) .
 \end{equation*}
Hence, the 2CTM reduces to the classical CTM \eqref{eq:sd_1}.
\end{rem}

\subsection{Sending and Receiving Functions for the 2CTM}
\begin{figure}[t]
\begin{center}
  \subfloat[]{\label{fig:ctm_sending}\includegraphics[width=0.4\textwidth]{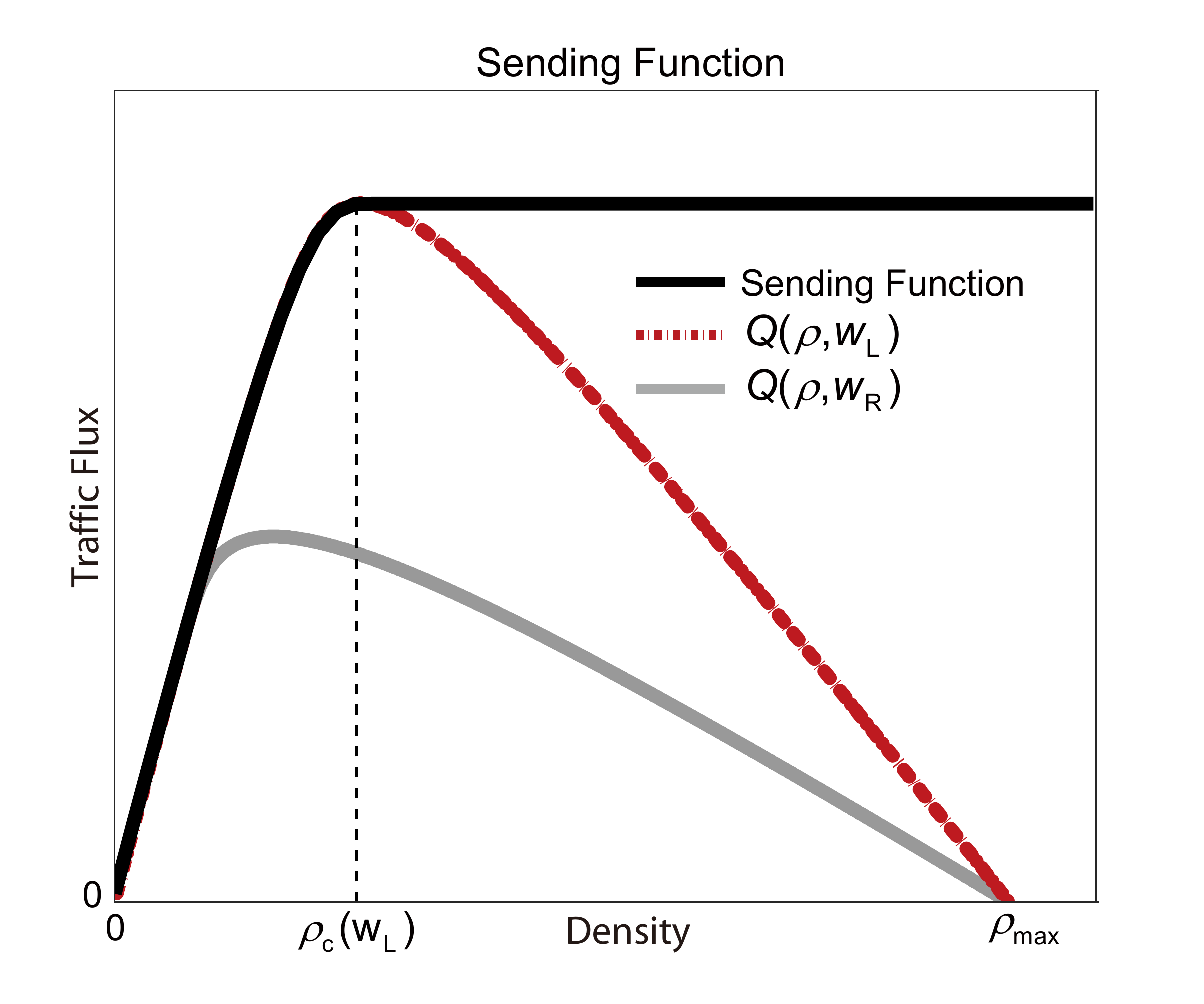}}
~\qquad
  \subfloat[]{\label{fig:ctm_receiving}\includegraphics[width=0.4\textwidth]{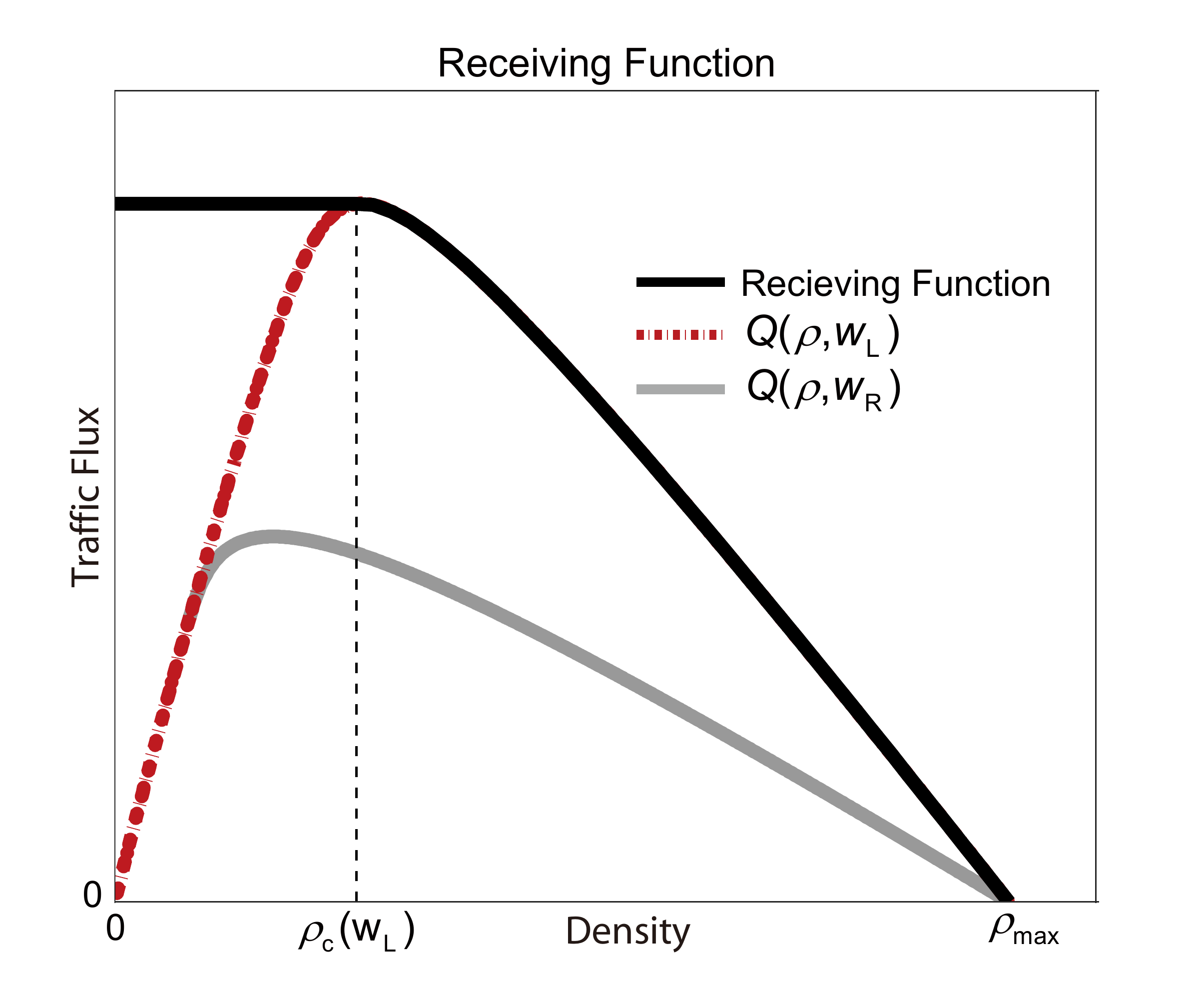}}
\caption{Sending (a) and receiving (b) functions of the 2CTM.}
\label{fig:sr}
\end{center}
\end{figure}

Let $u_{\text{L}}=\left(\rho_{\text{L}},\rho_{\text{L}}w_{\text{L}}\right)$ and $u_{\text{R}}=\left(\rho_{\text{R}},\rho_{\text{R}}w_{\text{R}}\right)$ be the traffic states of the upstream and downstream cells, respectively. The sending and receiving functions for the GSOM are analogous to the supply and demand functions proposed in \cite{lebacque2005second} (see Figure~\ref{fig:sr}). They have the form
\begin{equation}
 \label{eq:sr_old}
 \begin{split}
   S(\rho_{\text{L}},w_{\text{L}}) &= \left\{\begin{array}{rl} \rho_{\text{L}}v_{\text{L}} , &\text{if } \rho_{\text{L}}\le \rho_{\text{c}}(w_{\text{L}}) , \\
	         Q^{\text{max}}_{w_{\text{L}}} , &\text{if }  \rho_{\text{L}} > \rho_{\text{c}}(w_{\text{L}}) ,\end{array} \right.\quad \\[.4em]
   R(\rho_{\text{M}},w_{\text{L}}) &= \left\{\begin{array}{rl} Q^{\text{max}}_{w_{\text{L}}} , &\text{if }  \rho_{\text{M}} \le \rho_{\text{c}}(w_{\text{L}}) , \\
	           \rho_{\text{M}} v_{\text{M}} , &\text{if } \rho_{\text{M}}> \rho_{\text{c}}(w_{\text{L}}) ,\end{array} \right.
 \end{split}
\end{equation}
where $v_{\text{L}} = V(\rho_{\text{L}},w_{\text{L}})$ is the traffic velocity of the upstream vehicles, $Q^{\text{max}}_{w_{\text{L}}}$ is the capacity of the fundamental diagram $Q(\rho,w_{\text{L}})$, and $\rho_{\text{c}}(w_{\text{L}})$ represents the corresponding critical density. Here, the receiving function depends on an intermediate traffic state $u_{\text{M}}=(\rho_{\text{M}},\rho_{\text{M}}w_{\text{M}})$ \cite{AwRascle2000,lax1957hyperbolic}. This intermediate state is given as follows:
\begin{equation}
\label{eq:riemann}
\left\{\begin{array}{rl}
   &w_{\text{M}}  = w_{\text{L}} ,\\
  &v_{\text{M}} = V(\rho^*,w_{\text{M}}), \quad \text{where $\rho^*=\arg\min_{\rho}\left\{\left|v_R-V(\rho,w_{\text{M}})\right|\right\}$}\\
   &\rho_{\text{M}} , \quad \text{s.t. $v_{\text{M}} = V(\rho_{\text{M}},w_{\text{M}})$.}
\end{array} \right.
\end{equation}
Here $v_{\text{R}} = V(\rho_{\text{R}},w_{\text{R}})$ is the traffic velocity of the downstream vehicles. Equivalently, the middle state density $\rho_M$ can be computed as
\begin{equation*}
\rho_{M} = \text{argmin}_{\rho}\left\{V(\rho_{R},w_{R}) - V(\rho,w_{L}) \right\}.
\end{equation*}

\begin{figure}
\begin{center}
  \subfloat[]{\label{fig:intermediate_1}\includegraphics[width=0.4\textwidth]{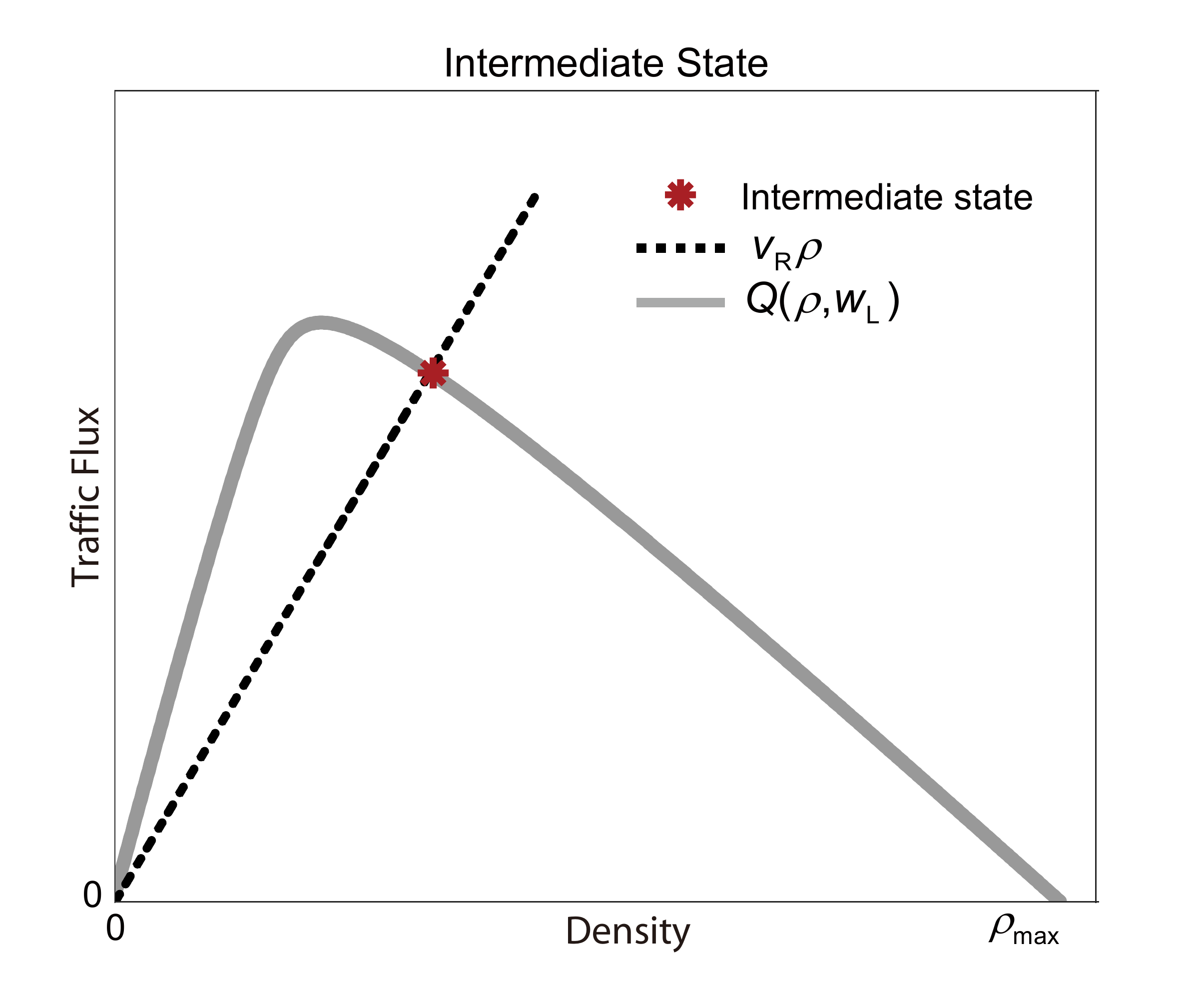}}
~\qquad
  \subfloat[]{\label{fig:intermediate_2}\includegraphics[width=0.4\textwidth]{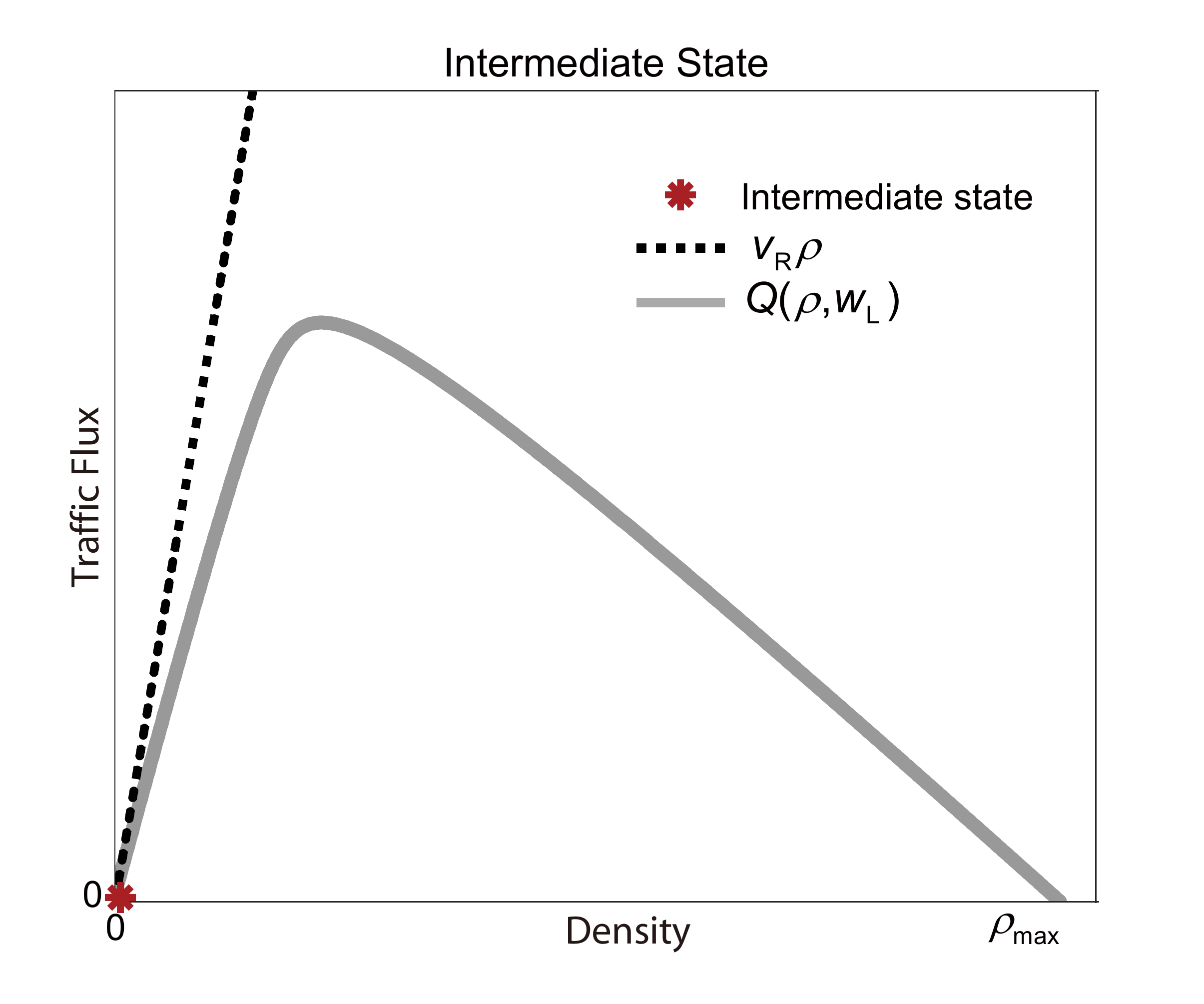}}
\caption{(a): Upstream vehicles have the potential to match the downstream velocity, i.e., $v_{\text{R}}\le V(0,w_{\text{L}})$. In this case, set $v_{\text{M}} = v_{\text{R}}$. (b): The maximum possible velocity of the upstream vehicles is lower than the downstream velocity, i.e., $v_{\text{R}}> V(0,w_{\text{L}})$. In this case, let $v_{\text{M}} = V(0,w_{\text{L}})$ in order to minimize the gap to the downstream velocity.}
\label{fig:intermediate_state}
\end{center}
\end{figure}

Note that in the case that upstream vehicles cannot match the downstream speed, i.e., $\max_{\rho}\{V(\rho,w_{\text{L}})\} = V(0,w_{\text{L}}) < v_{\text{R}}$ (the maximum possible velocity in the upstream cell is less than $v_{\text{R}}$), we let $v_{\text{M}} = V(0,w_{\text{L}})$. Otherwise, we always have $v_{\text{M}} = v_{\text{R}}$ (see Figure~\ref{fig:intermediate_state}). Thus, the middle state \eqref{eq:riemann} can be rewritten as:
\begin{equation}
\label{eq:riemann2}
\left\{\begin{array}{rl}
   &w_{\text{M}}  = w_{\text{L}} ,\\
  &v_{\text{M}} = V(0,w_{\text{L}}) ,\quad \text{ if }\quad V(0,w_{\text{L}}) < v_{\text{R}} ,\\
  &v_{\text{M}} = v_{\text{R}} , \quad \text{otherwise} ,\\
   &\rho_{\text{M}} , \quad \text{s.t. $v_{\text{M}} = V(\rho_{\text{M}},w_{\text{M}})$} .
\end{array} \right.
\end{equation}
The intuition behind this intermediate state is explained in more detail in the next subsection.

From \eqref{eq:sr_old} and \eqref{eq:riemann2}, it follows that the inflow into the $j$-th cell is given by
\begin{equation}
 \label{eq:flow_2}
    F^{\rho,n}_{j-1/2} = \min\left\{ S(\rho_{j-1}^{n},w_{j-1}^{n}) ,  R(\rho_{j-1/2}^{n},w_{j-1}^{n})\right\} ,
\end{equation}
where $\rho_{j-1/2}^{n}$ represents the intermediate density calculated from \eqref{eq:riemann2}, given the prior states $u^{n}_{j-1}$ and $u^{n}_{j}$. The outflow of the $j$-th cell, $F^{\rho,n}_{j+1/2}$, is defined analogously.

By comparing with the sending and receiving functions of the CTM \eqref{eq:ctm}, one can note the following: ($i$) the sending function in \eqref{eq:sr_old} is a function of $u_{\text{L}}$ only, like in the CTM; ($ii$) in contrast to the CTM, the receiving function now depends on the intermediate density $\rho_{\text{M}}$ and the upstream property $w_{\text{L}}$. The next subsection provides an explanation why the existence of the intermediate state, its dependence on the upstream property, as well as its influence on the receiving function, are in fact all physically reasonable.

\subsection{Interpretation of Intermediate Traffic State}
The existence of an intermediate state $u_{\text{M}}$ in the GSOM \eqref{eq:generic_conserv} can be understood as a consequence of the interactions of drivers with different properties $w$. Here, two adjacent cells (an upstream cell and a downstream cell) with initial states $u_{\text{L}}$ and $u_{\text{R}}$ are studied. The traffic flow through the cell interface is determined by the following rules:
\begin{enumerate}
 \item Downstream vehicles move out of the way (vehicles never move backwards). This creates space for the upstream vehicles to move in.

 \item Upstream vehicles maintain their properties when they transit from the upstream into the downstream cell, because the property $w$ is conserved for each vehicle.

 \item Vehicles from the upstream cell try to match their velocity to that of the downstream vehicles, to avoid collisions or falling behind. This means that $v_{\text{M}} = v_{\text{R}}$ whenever possible. If the incoming vehicles cannot achieve $v_{\text{R}}$, they assume their maximum possible speed, $v_{\text{M}} = V(0,w_{\text{L}})$.

 \item The incoming vehicles (with property $w_{\text{L}}$) adjust their spacing to achieve their new velocity $v_{\text{M}}$. This creates a new intermediate density $\rho_{\text{M}}$, selected so that $v_{\text{M}} = V(\rho_{\text{M}},w_{\text{L}})$.
\end{enumerate}
These rules provide a physical interpretation of the Riemann solver \eqref{eq:riemann2}.

From these rules, it becomes clear why the receiving capacity depends on the downstream velocity $v_{\text{R}}$ and the upstream property $w_{\text{L}}$, but not the downstream property $w_{\text{R}}$ or the upstream velocity $v_{\text{L}}$: it is the incoming drivers that are flowing through the interface; however, the new velocity that they assume depends on the space that the downstream vehicles have created, which in turn depends only on the downstream velocity.

Consider an example where the upstream drivers are quite passive, and the downstream cell is filled with aggressive drivers. When both the upstream and downstream cells have the same traffic density $\rho_{\text{L}} = \rho_{\text{R}}$, the amount of vehicles received by the downstream cell depends not only on the amount of the space that the downstream vehicles could generate (determined by $u_{\text{R}}$), but also on the willingness of the vehicles from the upstream cell to fill the free space (determined by $w_{\text{L}}$). Therefore, the receiving function \eqref{eq:sr_old} is also a function of the property of the upstream vehicles.

For the CGARZ model, the inverse function \eqref{eq:inverse_function_G}, defined in Section~\ref{section:inverse}, is applied, and the inflow of the $j$-th cell \eqref{eq:flow_2} becomes
\begin{equation*}
 \label{eq:flow_rho}
   \begin{split}
    F^{\rho,n}_{j-1/2} &= \min\left\{ S(\rho_{j-1}^{n},w_{j-1}^{n}), ~R(\rho_{j-1/2}^{n},w_{j-1}^{n})\right\} ,\\
    \rho_{j-1/2}^{n} &=\left\{\begin{array}{rl} G\left(V(0,w^{n}_{j-1}),w^{n}_{j-1}\right), &\text{if } V(0,w^{n}_{j-1})<V(\rho_j^n,w^{n}_{j}), \\
	G\left(V(\rho^{n}_{j},w^{n}_{j}),w^{n}_{j-1}\right), &\text{otherwise.} \end{array} \right.
    \end{split}
\end{equation*}%
The outflow can be defined in the same way.

\section{Model calibration and validation}\label{section:validation}
In this section, the CGARZ model is validated with field data, and is compared with a phase transition model as well as several second-order models that fit into the GSOM framework. The accuracy of a model is quantified by the deviation between the measurement data (the NGSIM trajectory data \cite{TrafficNGSIM} and the RTMC sensor data \cite{TrafficMnDOT}) and the various model predictions are obtained by applying the appropriate 2CTM\footnote{For the model validation on the NGSIM (resp.~RTMC) data, the discretization steps in the 2CTM are $\Delta t=0.05$s (resp.~$\Delta t=0.25$s) and $\Delta x=0.002$km (resp.~$\Delta x=0.008$km). These choices render the numerical approximation/discretization errors negligible relative to the model errors.} with initial and boundary conditions generated from the measurement data. 

\subsection{Considered Models and their Fundamental Diagrams}\label{sec:models_FD}
In this subsection, we list the traffic models compared in this study, and provide the formulas of the fundamental diagrams used in each model.

\subsubsection{The CGARZ Model}\label{sec:cgarz_models_FD}
In the CGARZ model, the fundamental diagram is given in \eqref{fd_cgarz}, where $Q_{\text{f}}(\rho)$ is provided in \eqref{eq:freediagram}, and $Q_{\text{c}}(\rho,w)$ reads:
 \begin{equation}
 \label{eq:congesteddiagram0}
 \begin{split}
   Q_{\text{c}}(\rho,w)&=Q_{\sigma(w),\mu(w)}\left(\rho\right)\\
   &=-\sigma(w) c \left(\left(\tfrac{\rho-\mu(w)}{\sigma(w)}\right)\tan^{-1}\!\left(\tfrac{\rho-\mu(w)}{\sigma(w)}\right)-\tfrac{1}{2}\ln\!\left(1+\left(\tfrac{\rho-\mu(w)}{\sigma(w)}\right)^2\right)\right)+ b \rho + C ,
 \end{split}
\end{equation}
where $\sigma(w)>0$, $\mu(w)>0$, and 
\begin{equation*}
 \label{eq:congesteddiagram1}
 \begin{split}
 &b=v_{\text{f}}-c\tan^{-1}\!\left(\tfrac{\rho_{\text{f}}-\mu\left(w\right)}{\sigma\left(w\right)}\right),\\
 &v_{\text{f}}=Q'_{\text{f}}\left(\rho_{\text{f}}\right),\\
 &c=-\frac{v_{\text{f}}\left(\rho_{\max}-\rho_{\text{f}}\right)+Q_{\text{f}}\left(\rho_{\text{f}}\right)}{\left(\tan^{-1}\left(\frac{\rho_{\text{f}}-\mu\left(w\right)}{\sigma\left(w\right)}\right)\right)\left(\rho_{\max}-\rho_{\text{f}}\right)+I}\\
  &I=\int_{\rho_{\text{f}}}^{\rho_{\max}}\tan^{-1}\!\left(\tfrac{\rho-\mu\left(w\right)}{\sigma\left(w\right)}\right)d\rho,
 \end{split}
\end{equation*}
and the constant $C$ is chosen such that $Q_c(\rho_{\text{f}})=Q_{\text{f}}\left(\rho_{\text{f}}\right)$. The reader is referred to \cite[Appendix B]{Fan2013} for details on the construction of the above flux function. 

Note that in \eqref{eq:congesteddiagram0}, $\sigma(w)$ and $\mu(w)$ are treated as functions of $w$, and their expressions are determined in the model calibration procedure based on the flow--density data (as detailed in Section~\ref{sec:w and parameters}). 

\subsubsection{The GARZ Model}\label{sec:garz_models_FD}
In the GARZ model, the physical interpretation of the property quantity $w$ is the empty-road velocity, i.e., $w:=V(0,w)$. The fundamental diagram is given as follows \cite[Chapter 3.1]{Fan2013}:
 \begin{equation}
 \label{eq:garzfd}
 \begin{split}
   Q(\rho,w)
   &=Q_{\alpha(w),\lambda(w),p(w)}(\rho)\\
   &=\alpha\left(w\right)\left(a_w+(b_w-a_w)\tfrac{\rho}{\rho_{\max}}-\sqrt{1+y_w^2}\right),
 \end{split}
\end{equation}
where $\alpha(w)>0$, $\lambda(w)>0$, $0<p(w)<1$ and
 \begin{equation*}
 \label{eq:garzfd_aby}
 a_w=\sqrt{1+\left(\lambda\left(w\right)p\left(w\right)\right)^2},\quad b_w=\sqrt{1+\left(\lambda\left(w\right)\left(1-p\left(w\right)\right)\right)^2},
\end{equation*}
and
\begin{equation}
y_w=\lambda\left(w\right)\left(\tfrac{\rho}{\rho_{\max}}-p\left(w\right)\right).
\end{equation}
Similarly to the CGARZ model, $\alpha(w)$, $\lambda(w)$, and $p(w)$ are treated as functions of $w$ and are calibrated from the flow--density data.
\begin{rem}
In the GARZ model, the property $w$ is chosen as the empty-road velocity. In the CGARZ model this is not possible, because the collapsing no longer distinguishes different flow--density or velocity--density curves near the vacuum ($\rho=0$).
\end{rem}

\subsubsection{The ARZ Model}\label{sec:arz_models_FD}
Like the GARZ model, the property quantity in the ARZ model is defined as the empty-road velocity, $w:=V(0,w)$. As described in \eqref{eq:velocity_arz}, the velocity function $V(\rho,w)$ is obtained by shifting $w-V_{\text{eq}}(0)$ from the equilibrium velocity function $V_{\text{eq}}(\rho)$. In the remainder, let $V_{\text{eq}}(\rho)$ be defined as
\begin{equation*}
\label{eq:arz_eq_v}
\begin{split}
V_{\text{eq}}\left(\rho\right)=\frac{Q\left(\rho, w_{\text{eq}}\right)}{\rho},
\end{split}
\end{equation*}
with $Q(\rho, w_{\text{eq}})$ computed according to \eqref{eq:garzfd}. Given the above definition, the ARZ and GARZ models studied in this section share the same equilibrium velocity function and have the same equilibrium FD. The reader is referred to 
\cite[Chapter 3.2]{fan2014comparative} for an illustration of the flow--density curves of the ARZ model.

\subsubsection{The LWR Model}\label{sec:lwr_models_FD}
The fundamental diagram of the LWR model is chosen to be identical to the equilibrium FD of the GARZ model, i.e., $Q(\rho)=Q(\rho, w_{\text{eq}})$ with $Q(\rho, w_{\text{eq}})$ computed according to \eqref{eq:garzfd}.

\subsubsection{The PT Model}\label{sec:gpt_models_FD} 
The PT model is defined using different dynamics in the free-flow regime  and the congestion regime:
\begin{equation*}\label{eq:ptfd}
\left\{\begin{array}{ll}
	        \rho_t + \left(\rho V_{\text{f}}\left(\rho\right)\right)_x = 0, &\quad\text{in free-flow ($\Omega_{\text{f}}$)},\\
	         \left\{\begin{array}{ll}
	        \rho_t + \left(\rho V_{\text{c}}\left(\rho,w\right)\right)_x = 0, \\
	         y_t + \left(y V_{\text{c}}\left(\rho,w\right)\right)_x = 0,\end{array} \right. &\quad\text{in congestion ($\Omega_{\text{c}}$)},\end{array} \right. 
\end{equation*}
where $y=\rho w$, $\Omega_{\text{f}}$ and $\Omega_{\text{c}}$ are the respective domains of validity of the free-flow and congested equations of the model defined explicitly below. The velocity functions of the PT model implemented in this work for the free-flow regime and for the congestion regime are given as:
\begin{equation*}\label{eq:ptfdV}
 V_{\text{f}}(\rho) = \frac{Q_{\text{f}}(\rho)}{\rho} = v_\text{max}\left(1-\frac{\rho}{\tilde{\rho}_\text{max}}\right),\quad V_{\text{c}}(\rho,w) = \frac{Q_{\text{c}}(\rho,w)}{\rho} =  \frac{w}{\rho}\left(\frac{\rho_{\max}-\rho}{\rho_{\max}-\rho_c}\right).
\end{equation*}
Given the above definition of the velocity functions, the free-flow and the congested regimes are defined as follows:
\begin{equation*}\label{eq:ptfdDOMAIN}
\left\{\begin{array}{ll}
\Omega_{\text{f}}=\left\{\left(\rho,w\right)\left|\enskip 0\le \rho \le \rho_{\max},\enskip V_{\text{f}}\left(\rho\right) \ge v_{\text{f},\min}\right.\right\},\\
\Omega_{\text{c}}=\left\{\left(\rho,w\right)\left|\enskip 0\le\rho \le \rho_{\max},\enskip V_{\text{c}}\left(\rho,w\right) \le v_{\text{c},\max},\enskip w_{\min} \le w \le w_{\max}\right.\right\},\\\end{array} \right. 
\end{equation*}
where $v_{\text{f},\min}$ is the minimal velocity in free--flow, and $v_{\text{c},\max}$ is the maximal velocity in congestion. The lower and upper bounds of $w$ in the congestion regime are denoted as $w_{\min}$ and $w_{\max}$, respectively.



\subsection{Data-Fitting for Macroscopic Traffic Models}\label{sec:data_fitted}

In this subsection, we present the individual steps in the traffic model calibration procedure, which is a \textit{weighted least squares} (WLSQ) approach. The main challenge to fit models in the GSOM family is due to the fact that a whole family of fundamental diagrams must be determined (rather than a single curve). Note that the general flux function is written as a function of the density and the property, i.e., $Q(\rho,w)$. An alternative view which is useful for calibration is to recognize that the property $w$ only influences the parameters of the traditional flow-density fundamental diagram (e.g., see \eqref{eq:congesteddiagram0}--\eqref{eq:garzfd} where $w$ only appears in the parameters $\alpha,\lambda,$ and $p$). Unfortunately, the functional relationship between the  flow-density curve parameters and the property variable is not known.

We use the above points to design a GSOM calibration procedure that occurs in three steps. In Step~1, we calibrate all parameters associated with the equilibrium fundamental diagram. The second and third steps are required to empirically determine the flow--density parameter dependence on the property. In Step 2, we use a weighted least squares calibration procedure to fit a family of fundamental diagrams for a set of discrete values of the property, with one set of optimal parameters associated with each property value.  Finally in Step 3, we apply a polynomial regression on the optimal parameters for the given values of the property determined in Step 2, to construct a best fit between the property value and the property-dependent flow-density curve parameters.

We first present the particular WLSQ data-fitting approach \cite{fan2014comparative} that is used for calibration. Let $(\rho_j, Q_j)$ represent a flow--density data pair, and denote by $n_{\text{data}}$ the total number of data pairs. In addition, denote as $\beta\in(0, 1)$ the weight parameter of the WLSQ, and let $\theta_{\beta}$ be the set of parameters to be calibrated which is parameterized by $\beta$. Given a parameter set $\theta_{\beta}$, the predicted flux corresponding to density measurement $\rho_j$ is given by $Q_{\theta_{\beta}}(\rho_j)$. In general, $Q_{\theta_{\beta}}(\rho_j)-Q_j\neq 0$ (i.e., the predicted flux corresponding to the measured density does not match the measured flux), and we can penalize the discrepancy in a cost functional to determine the best parameters $\theta_{\beta}^{*}$.

In the WLSQ algorithm, the following minimization problem is solved to determine the optimal parameter set:
\begin{equation}
\label{eq:WLSQ_f}
\theta^{*}_{\beta}=\arg\min_{\theta_{\beta}} F_{\beta}\left(\theta_{\beta}\right),
\end{equation}
where the cost functional $F_{\beta}\left(\theta\right)$ is defined as:
\begin{equation}
\label{eq:WLSQ_prelim}
F_{\beta}\left(\theta\right)=
\beta \sum^{n_{\text{data}}}_{j=1}{\left((Q_{\theta}(\rho_{j})-Q_{j})_{+}\right)^2}+
\left(1-\beta\right) \sum^{n_{\text{data}}}_{j=1}\left((Q_{\theta}(\rho_{j})-Q_{j})_{-}\right)^2,
\end{equation}
with
\begin{align*}
\left(Q_{\theta}(\rho_{j})-Q_{j}\right)_{+}
&= \max\left\{ Q_{\theta}(\rho_{j})-Q_{j},0\right\} , \\
\left(Q_{\theta}(\rho_{j})-Q_{j}\right)_{-}
&= \max\left\{-Q_{\theta}(\rho_{j})+Q_{j},0\right\} .
\end{align*}
Hence, the WLSQ minimizing parameters depend on the weighting parameter $\beta$. For large $\beta$, data below the curve $Q_{\theta}(\rho)$ is strongly penalized, thus the minimization problem \eqref{eq:WLSQ_f} tends to let more data points stay above the calibrated curve. Hence, when $\beta \rightarrow 1$, almost all the data points are above the calibrated $Q_{\theta}(\rho)$, and the reverse is true when $\beta \rightarrow 0$. Note that \eqref{eq:WLSQ_f} reverts to the classical \textit{least squares} (LSQ) fit when $\beta=0.5$, which corresponds to the calibration of the equilibrium curve.  


\subsubsection{Step~1: Calibrating the Equilibrium Fundamental Diagram}
We briefly describe the process to determine the equilibrium fundamental diagram. The equilibrium fundamental diagram includes both parameters that depend on the property, as well as parameters that are independent of the property.  For example, in the CGARZ model, the property--invarient parameters calibrated in this step include  $v_{\max}$, $\rho_{\text{f}}$, $\tilde{\rho}_{\max}$, and $\rho_{\max}$, while the parameters that depend on the property include $\sigma$ and $\mu$. The property--dependent parameters are determined at the equilibrium value of the property,  $\sigma_{\text{eq}}$ and $\mu_{\text{eq}}$. Let $\theta_{\text{eq}}$ denote the parameters that are calibrated in Step~1 of the calibration procedure. Table~\ref{tb:parameters} lists the parameters in $\theta_{\text{eq}}$ for all the studied models (in the column named ``Step 1''). 

\renewcommand{\arraystretch}{1.5}%
\begin{table}
  \centering
  \begin{threeparttable}[b]
  \begin{tabular}{c|c|c|c}
  \toprule
  \multirow{2}{*}{Model} &\multicolumn{3}{c}{Parameters calibrated in}\\
  & \multicolumn{1}{c}{Step~1} & \multicolumn{1}{c}{Step~2} & \multicolumn{1}{c}{Step~3}\\ \hline\hline
  \multirow{2}{*}{CGARZ}
  & $\theta_{\text{eq}}=$\{$v_{\max}$, $\rho_{\text{f}}$,
  & $\theta_{\beta_{i}}=$\{$\sigma_{\beta_i}$, $\mu_{\beta_i}$\} & $a_{\ell,\sigma}$, $a_{\ell,\mu}$\\[-1ex]
  & $\tilde{\rho}_{\max}$, $\rho_{\max}$, $\sigma_{\text{eq}}$, $\mu_{\text{eq}}$\} & for $i\in \{1,\cdots,m\}$ & for $\ell \in \{0,1,\cdots,k\}$\\
  \hline
  \multirow{2}{*}{GARZ}  & $\theta_{\text{eq}}=$\{$\rho_{\max}$, & $\theta_{\beta_{i}}=$\{$\alpha_{\beta_i}$, $\lambda_{\beta_i}$, $p_{\beta_i}$\} & $a_{\ell,\alpha}$, $a_{\ell,\lambda}$, $a_{\ell,p}$\\[-1ex]
  & $\alpha_{\text{eq}}$, $\lambda_{\text{eq}}$, $p_{\text{eq}}$\} & for $i \in \{1,\cdots,m\}$ & for $\ell \in \{0,1,\cdots,k\}$\\
  \hline
  ARZ & Same as GARZ & None & None\\
  \hline
  LWR & Same as GARZ & None & None\\
  \hline
  \multirow{2}{*}{PT}
  & $\theta_{\text{eq}}=$\{$v_{\max}$, $\tilde{\rho}_{\max}$,
  & $\theta_{\beta_{i}}=$\{$w_{\beta_i}$\}
  &\multirow{2}{*}{None}\\[-1ex]
  & $\rho_{\max}$, $w_{\text{eq}}$\} &for $i \in \{1,\cdots,m\}$&\\
  \bottomrule
  \end{tabular}
  \end{threeparttable}
  \vskip.1em
\caption{Parameters to be calibrated in Step~1, Step~2 and Step~3 of the calibration procedure for all the considered models.}
 \label{tb:parameters}
\end{table}

\vspace{1em}
\noindent \textit{Procedure for the CGARZ model.} In order to determine the free-flow threshold density $\rho_{\text{f}}$, below which all the curves collapse, several flow--density curves in the family of FDs need to be calibrated simultaneously with the equilibrium FD. For the $i$-th curve calibrated together with the equilibrium FD, its weight parameter is defined as $\beta_{i,\text{eq}}$, and its parameter set to be calibrated is defined as $\theta_{\beta_{i,\text{eq}}}=\{v_{\max}, \rho_{\text{f}}, \tilde{\rho}_{\max}, \rho_{\max},\sigma_{\beta_{i,\text{eq}}}, \mu_{\beta_{i,\text{eq}}}\}$ (where the first four parameters are the shared parameters with the equilibrium FD). To calibrate a total of $m_{\text{eq}}$ curves together with the equilibrium FD, the following minimization problem is solved:
\begin{equation}
\label{eq:WLSQ_together}
\min_{\mathcal{Q}} \left\{\tilde{F}_{0.5}\left(\theta_{\text{eq}}\right)+\sum_{i=1}^{m_{\text{eq}}}\tilde{F}_{\beta_{i,\text{eq}}}\left(\theta_{\beta_{i,\text{eq}}}\right)\right\},
\end{equation}
where $\mathcal{Q}=(\bigcup_{i=1}^{m_{\text{eq}}} \theta_{\beta_{i,\text{eq}}})\bigcup \theta_{\text{eq}}$, and $\tilde{F}_{\beta}(\theta)$ is a modification of $F_{\beta}(\theta)$ in \eqref{eq:WLSQ_prelim}:
\begin{equation*}
\label{eq:WLSQ_shrink}
\tilde{F}_{\beta}\left(\theta\right)=
\beta\sum^{n_{\text{data}}}_{j=1}{\left(S^{\tau}_{\rho_{\text{f}}}\left(\left(Q_{\theta}(\rho_{j})-Q_{j}\right)_{+}\right)\right)^2}+
\left(1-\beta\right) \sum^{n_{\text{data}}}_{j=1}{\left(S^{\tau}_{\rho_{\text{f}}}\left(\left(Q_{\theta}(\rho_{j})-Q_{j}\right)_{-}\right)\right)^2}.
\end{equation*}
Here $\tau>0$, and the \textit{shrinkage operator} is given by $S^{\tau}_{\rho_{\text{f}}}$ via
\begin{equation*}
S^{\tau}_{\rho_{\text{f}}}\left(\left(Q_{\theta}(\rho_{j})-Q_{j}\right)_{\pm}\right) = 	\left\{\begin{array}{ll}
	        0, &\textrm{if $\rho_j<\rho_{\text{f}}$ and $\left(Q_{\theta}(\rho_{j})-Q_{j}\right)_{\pm}<\tau$},\\
	         \left(Q_{\theta}(\rho_{j})-Q_{j}\right)_{\pm}, &\textrm{otherwise.}\end{array} \right.
\end{equation*}

Shrinkage methods (similarly $\epsilon$--insensitive bands) are commonly used  to promote sparsity in the decision variables in optimization and regression problems \cite{vapnik1995nature,tibshirani1996regression}. For example they are used to avoid fitting noise in regression problems. Here we use it similarly to avoid using a family of curves to represent the flow density relationship when a single curve is sufficient. Without a shrinkage operator, the calibrated density $\rho_{\text{f}}$ where all the curves collapse would be very small (e.g., $\rho_{\text{f}}<10^{-5}$ veh/km/lane), because the flow--density data points in the low density regime do have a spread (although much less than in the high density domain). Due to the shrinkage operator in the functional, the optimization problem \eqref{eq:WLSQ_together} promotes the collapsing of the data-fitted curves until a transition point in the observed data when the spread becomes larger, e.g., around $\rho_{\text{f}}\in [50,80]$ veh/km/lane.

\begin{figure}
\begin{center}
  \subfloat[]{\label{fig:fig_cgarz_eq_cali_ngsim_c}\includegraphics[width=0.4\textwidth]{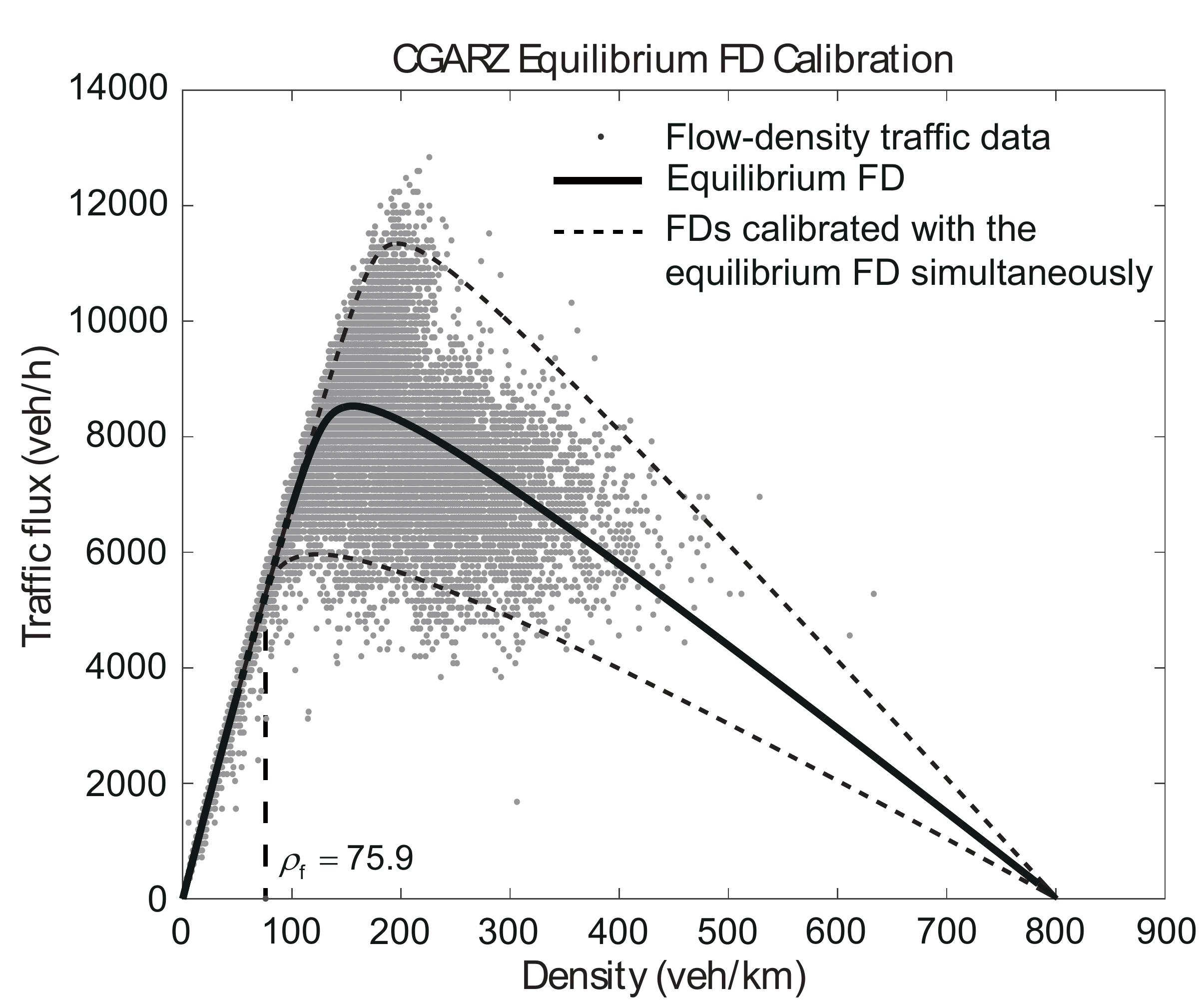}}
~\qquad
  \subfloat[]{\label{fig:fig_cgarz_cali_ngsim_c}\includegraphics[width=0.4\textwidth]{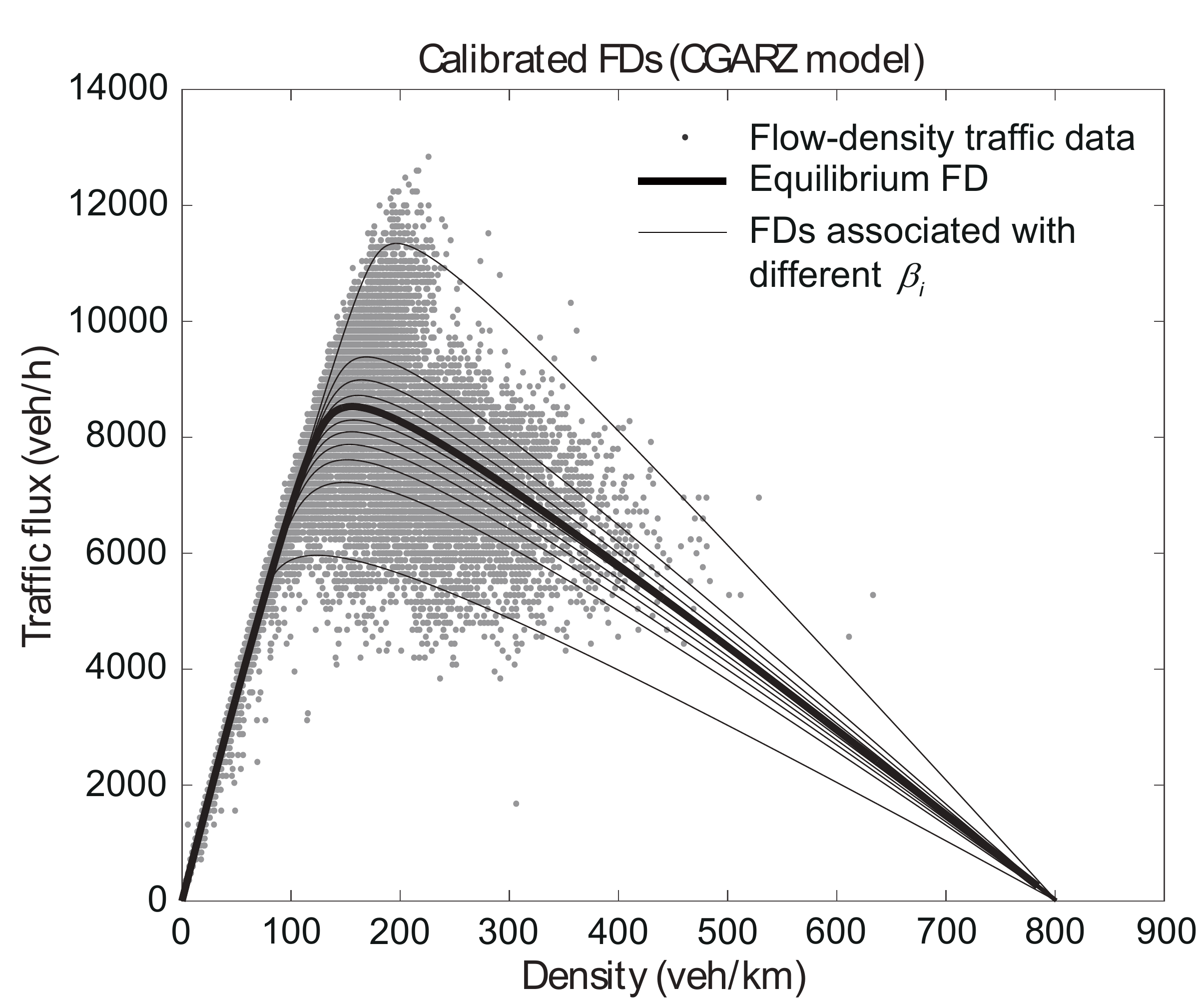}}
\caption{Calibration of the CGARZ model using the flow--density data provided by the PeMS loop detectors near the freeway segment that generates the NGSIM dataset. The calibrated parameters are used in the model validations on the NGSIM dataset. (a): The calibrated equilibrium FD of the CGARZ model and the two FDs (corresponding to $\beta_{1,\text{eq}}=0.001$ and $\beta_{2,\text{eq}}=0.999$) calibrated together with the equilibrium FD. (b): The calibrated family of FDs shown for a selection of $\beta$-values (i.e., $\beta_i$ in \eqref{eq:WLSQ_beta_family} with $i$ given in \eqref{eq:WLSQ_beta_family_plot}). Note that the family of FD curves is defined for any real $\beta\in [0,1]$ rather than just the choices plotted as curves, which is also true in Figure \ref{fig:fig_ngsim_cali_c} and Figure \ref{fig:fig_minn_cali_c}. }
\label{fig:fig_cgarz_rtmc_cali}
\end{center}
\end{figure}

As an example, Figure~\ref{fig:fig_cgarz_eq_cali_ngsim_c} illustrates the calibration of the CGARZ equilibrium FD using the flow--density traffic data provided by the loop detectors\footnote{The loop detector data is obtained from the \textit{Freeway Performance Measurement System} (PeMS)\cite{TrafficPEMS}.} near the freeway segment that generates the NGSIM dataset. In Figure~\ref{fig:fig_cgarz_eq_cali_ngsim_c}, two FD curves are calibrated together with the equilibrium FD, i.e., $m_{\text{eq}}=2$, with $\beta_{1,\text{eq}}=0.001$, $\beta_{2,\text{eq}}=0.999$, and $\tau=400$ veh/hr/lane. We also conduct a sensitivity analysis of the shrinkage operator to investigate the dependence of the calibrated FDs on different $\tau$-values. The outcome is that the obtained FDs are rather insensitive to $\tau$. For example, the calibrated $\rho_{\text{f}}$ varies only in a small range between 70 and 80 veh/km/lane for all $300\le\tau\le 700$ veh/hr/lane. 

\vspace{1em}
\noindent \textit{Procedure for the other models.} For the GARZ and the PT model, the equilibrium FD can be calibrated without considering any non-equilibrium curves. Letting $\beta=0.5$ in \eqref{eq:WLSQ_f} and \eqref{eq:WLSQ_prelim}, we obtain the classical LSQ fitting problem:
\begin{equation}
\label{eq:LSQ_0}
\min_{\theta_{\text{eq}}}  \sum_{j=1}^{n_{\text{data}}} \left(Q_{\theta_{\text{eq}}}(\rho_{j})-Q_{j}\right)^2,
\end{equation}
which is used to determine the parameters in the equilibrium curve.

Note that as stated in Section~\ref{sec:arz_models_FD} and Section~\ref{sec:lwr_models_FD}, the equilibrium FD of the ARZ model and the FD of the LWR model are the same as the equilibrium FD of the GARZ model. Hence, the parameters calibrated for the equilibrium FD of the GARZ can be directly applied to the ARZ and LWR models.

\subsubsection{Step~2: Calibrating a Family of Fundamental Diagrams}
In Step~2, we calibrate a family of FDs using the WLSQ fitting (i.e., each calibrated FD is associated with a weight parameter value $\beta$ and a property $w$), so that we can later determine a relationship between the property and the property-dependent parameters via regression in Step 3. The FD parameters that are invariant with respect to $\beta$ (i.e., $v_{\max}$, $\rho_{\text{f}}$, $\tilde{\rho}_{\max}$, $\rho_{\max}$ in the CGARZ model; $\rho_{\max}$ in the GARZ model; and $v_{\max}$, $\tilde{\rho}_{\max}$, $\rho_{\max}$ in the PT model) have already been determined in Step~1 and are now treated as fixed. Hence, it remains to calibrate the FD parameters whose values depend on $\beta$. In Step~2, we calibrate these FD parameters for a set of fixed values of $\beta$ (i.e., $\beta\in\mathcal{B}=\{\beta_1,\cdots,\beta_{i},\cdots,\beta_{m}\}$). Denote as $\theta_{\beta_i}$ the parameter set associated with $\beta=\beta_i$. For each model, the parameters to be determined in Step~2 of the calibration procedure are listed in Table~\ref{tb:parameters}. The WLSQ fitting is applied to determine the parameters associated with each $\beta\in\mathcal{B}$:
\begin{equation}
\label{eq:WLSQ_beta}
\theta^{*}_{\beta_i}=\arg\min_{\theta_{\beta_i}} F_{\beta_i}\left(\theta_{\beta_i}\right), \quad \textrm{for all $i\in\{1,\cdots,m\}$,}
\end{equation}
where $F_{\beta}(\theta)$ is given in \eqref{eq:WLSQ_prelim}. Following the example in Figure~\ref{fig:fig_cgarz_eq_cali_ngsim_c}, we calibrate a family of FDs associated with various $\beta \in \mathcal{B}$ where
\begin{equation}
\label{eq:WLSQ_beta_family}
\beta_i=0.001+\frac{0.998}{99}(i-1).\quad i\in\{1,\cdots,100\}.
\end{equation}
For better visibility, we only plot in Figure~\ref{fig:fig_cgarz_cali_ngsim_c} the calibrated curves associated with $\beta_i$ where
\begin{equation}
\label{eq:WLSQ_beta_family_plot}
i\in\{1,11,21,31,41,51,61,71,81,91,100\}.
\end{equation}


\subsubsection{Step~3: Determining the Relationship between Property and Parameters}\label{sec:w and parameters}
In Step~3, the expressions of $\sigma(w)$, $\mu(w)$ in the CGARZ model, and the expressions of $\alpha(w)$, $\lambda(w)$, $p(w)$ in the GARZ model are determined using a polynomial regression. 

We take $\sigma(w)$ in the CGARZ as an example to illustrate the polynomial regression procedure, and note the same approach applies to other property--dependent parameters. The function $\sigma(w)$ is modeled as an $k$-th degree polynomial:
\begin{equation*}
\label{eq:polyfit}
\sigma(w)=\sum_{l=0}^ka_{l,\sigma}w^l.
\end{equation*}
Denoting as $a_{\sigma}=(a_{0,\sigma},\cdots,a_{k,\sigma})$, a polynomial regression is used to determine $a_{\sigma}$. For each $\sigma_{\beta_i}$ obtained in Step~2, we compute its associated property value $w_{\beta_i}$, which (in the model validation of this article\footnote{The property value of the CGARZ model can be assigned with other physical meanings as well, e.g., the fraction of flow represented by smart vehicles, as long as $\beta$ and $w$ are related in an invertible function $w(\beta)$.}) is the maximum flow of the FD curve associated with the weight parameter $\beta_i$. Hence, the obtained $(\sigma_{\beta_i},w_{\beta_i})$ pairs for all $i\in\{1,\cdots,m\}$ can be used in the polynomial regression. The parameters for the CGARZ and the GARZ models to be determined in Step~3 are listed in Table~\ref{tb:parameters}.

More details about the implementation of the calibration are available in the supplementary source code \url{https://github.com/Lab-Work/GSOM_LWR_PT}.

\subsection{Model Validation with the NGSIM Sensor Data}\label{sec:validation_ngsim}

\subsubsection{Analysis Setup}\label{sec:setup_ngsim}
As in \cite{fan2014comparative,FanSeibold2013}, we consider the NGSIM dataset \cite{TrafficNGSIM}, collected on 04/13/2005 on a 500 meter six-lane segment of the eastbound direction of I-80 located in Emeryville, CA. Using video cameras, the precise trajectories of all vehicles in the segment are accessible (with a temporal resolution of 0.1 seconds), in three 15-minute intervals at the onset during rush hour: 4:00pm--4:15pm, 5:00pm--5:15pm, and 5:15pm--5:30pm. To construct traffic density and velocity at location $x\in[0,L]$ and time $t\in[0,T]$ from trajectory data, the following \textit{kernel density estimation} technique \cite{Parzen1962, Rosenblatt1956} is applied:
\begin{equation}
\label{eq:kernel_density_estimation}
\begin{split}
\rho^{\text{data}}(x,t) &= \sum_{j=1}^{n_{\text{tra}}} K(x-x_j(t))\;,\\
Q^{\text{data}}(x,t) &= \sum_{j=1}^{n_{\text{tra}}} v_j K(x-x_j(t))\;,\\ v^{\text{data}}(x,t) &= \frac{Q^{\text{data}}(x,t) }{\rho^{\text{data}}(x,t)}\;,
\end{split}
\end{equation}
where $K(x) = \tfrac{1}{\sqrt{2\pi}h}e^{-\frac{x^2}{2h^2}}$ is a Gaussian kernel, $n_{\text{tra}}$ is the total number of trajectories, and $x_j$ and $v_j$ represent the position and velocity data of the $j$-th vehicle, respectively. In this analysis, we set $h=25$m. The boundary conditions for the studied freeway segment are $\rho(0,t)$, $v(0,t)$, $\rho(L,t)$, and $v(L,t)$. The initial conditions are the traffic state at $t=0$, i.e., $\rho(x,0)$ and $v(x,0)$ for $x\in[0,L]$. In order to study the prediction accuracies of the considered traffic models, we compare the traffic states they predict (based on the aforementioned initial and boundary conditions) with the corresponding traffic states \eqref{eq:kernel_density_estimation} constructed from the trajectory data.

\begin{figure}
\begin{center}
  \subfloat[]{\label{fig:fig_garz_cali_ngsim_c}\includegraphics[width=0.4\textwidth]{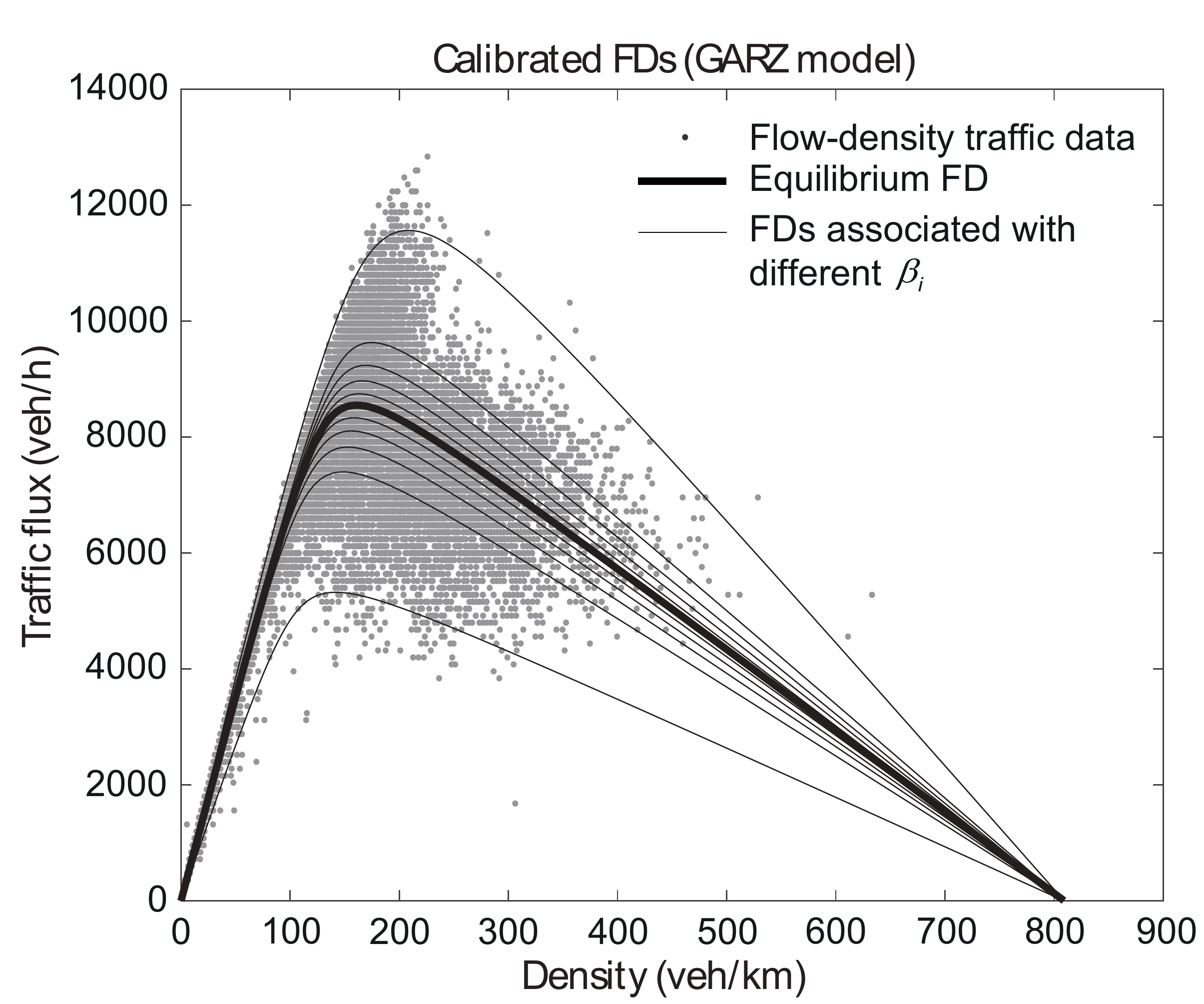}}
~\qquad
  \subfloat[]{\label{fig:fig_pt_cali_ngsim_c}\includegraphics[width=0.4\textwidth]{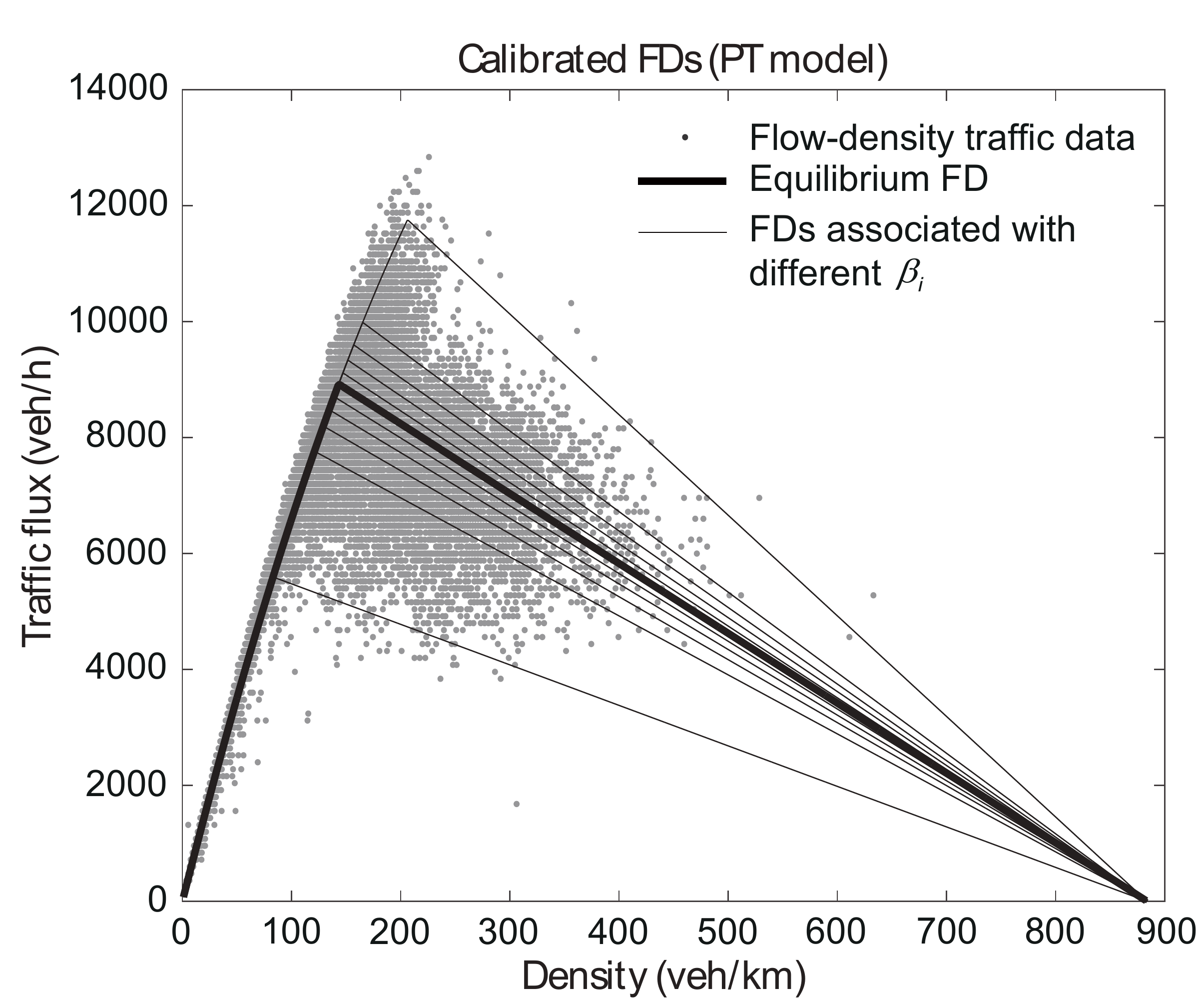}}
\caption{Model calibration using the flow--density data provided by the PeMS loop detectors near the freeway segment generating the NGSIM dataset, and the calibrated parameters are used in the model validations on the NGSIM dataset. (a-b): The calibrated family of fundamental diagrams of the GARZ (a) and the PT (b) models. The calibrated family of FDs are plotted for a selection of $\beta$-values (i.e., $\beta_i$ in \eqref{eq:WLSQ_beta_family} with $i$ given in \eqref{eq:WLSQ_beta_family_plot}).}
\label{fig:fig_ngsim_cali_c}
\end{center}
\end{figure}

The parameters in the traffic models are calibrated based on the historic flow--density data near the considered freeway segment, which are obtained via loop detectors and are provided by the PeMS \cite{TrafficPEMS}. The calibrated parameters for all the traffic models are detailed in Table~\ref{tb:parameters_pems} in the Appendix. To obtain the free-flow threshold $\rho_{\text{f}}$ for the CGARZ model in Step~1, we calibrate two curves together with the equilibrium curve (as shown in Figure~\ref{fig:fig_cgarz_eq_cali_ngsim_c}). In Step~2, a family of 100 curves is calibrated for the CGARZ (Figure~\ref{fig:fig_cgarz_cali_ngsim_c}), GARZ (Figure~\ref{fig:fig_garz_cali_ngsim_c}), and the PT models (Figure~\ref{fig:fig_pt_cali_ngsim_c}), where the family of $\beta_{i}$ is given in \eqref{eq:WLSQ_beta_family}\footnote{For better visibility, we only plot the curves associated with $\beta_i$ for $i$ given in \eqref{eq:WLSQ_beta_family_plot}.}. In Step~3, a polynomial regression is applied to construct the relationship between $w$ and the obtained parameters.

\subsubsection{Error Metric}\label{sec:error_metric_ngsim}
For the model validation using the NGSIM data, we consider the spatio-temporal average errors
\begin{equation*}
 \label{eq:error_integral_ngsim}
E_{\rho} = \frac{1}{TL}\int_0^T\int_0^L E_{\rho}(x,t)\ud{x}\ud{t}\;,\qquad
E_{v} = \frac{1}{TL}\int_0^T\int_0^L E_{v}(x,t)\ud{x}\ud{t}\;,
\end{equation*}
where
\begin{equation*}
\label{eq:error_measure_x}
E_{\rho}(x,t) = \frac{\left|\rho^\text{model}(x,t)-\rho^\text{data}(x,t)\right|}{n_{\text{lane}}} ,\quad E_{v}(x,t) = \left|v^\text{model}(x,t)-v^\text{data}(x,t)\right|.
\end{equation*}
Here, $\rho^\text{model}(x,t)$ and $v^\text{model}(x_{s2},t)$ are the predicted density and velocity, respectively, at location $x$ and time $t$, and $\rho^\text{data}(x,t)$ and $v^\text{data}(x_{s2},t)$ are obtained according to \eqref{eq:kernel_density_estimation}.

\subsubsection{Model Validation Results}\label{sec:results_ngsim}

{\renewcommand{\arraystretch}{1.5}%
\begin{table}
  \centering
  \begin{threeparttable}[b]
  \begin{tabular}{c|c|ccccc}
  \toprule
  Dataset & & CGARZ & GARZ & ARZ & LWR & PT \\\hline
  {NGSIM} & $E_{\rho}$ & 6.96(+5.8\%) & \textbf{6.58} & 7.31(+11.1\%) & 7.90(+20.1\%) & 6.98(+6.1\%)\\[-1ex]
   4:00pm-4:15pm& $E_v$ & 4.52(+4.6\%)& \textbf{4.32} & 4.73(+9.5\%) & 6.19(+43.3\%) & 4.56(+5.6\%)\\\hline
  {NGSIM} & $E_{\rho}$ & 6.85(+2.1\%) & \textbf{6.71} & 9.24(+37.7\%) & 8.05(+20.0\%) & 7.46(+11.2\%)\\[-1ex]
    5:00pm-5:15pm& $E_v$ & 3.84(+4.3\%) & \textbf{3.68} & 4.68(+27.2\%) & 4.72(+28.3\%) & 4.00(+8.7\%)\\\hline
   {NGSIM} & $E_{\rho}$ & \textbf{7.54}& 7.56(+0.3\%)& 12.84(+70.3\%) & 8.39(+11.3\%) & 8.93(+18.4\%) \\[-1ex]
   5:15pm-5:30pm & $E_v$ & \textbf{3.51} & 3.61(+2.8\%) & 5.63(+60.4\%) & 4.54(+29.3\%) & 4.04(+15.1\%) \\\bottomrule
  \end{tabular}
  \end{threeparttable}
  \vskip.4cm
\caption{Density prediction error $E_{\rho}$ and velocity prediction error $E_v$ given by the considered models on the three time intervals of the NGSIM dataset. The bold numbers are the minimum errors in each row, and the percentage in the parenthesis next to each error is the relative increase of the error compared to the minimum error in the same row.}
 \label{tb:ngsim_result}
\end{table}

The average prediction errors $E_{\rho}$ and $E_v$ of all the traffic models are reported in Table~\ref{tb:ngsim_result}. Based on the presented results, the key findings are as follows:
\begin{enumerate}[1.]
\item Due to the high congestion level in the studied location and time period (see the data points in Figure~\ref{fig:fig_ngsim_cali_c}), the prediction accuracy of a given traffic model is mainly determined by its capability of modeling the spread of the traffic data in the congested regime. Hence, the CGARZ and the GARZ models provide the most accurate density and velocity predictions.

\item A closer look at the NGSIM data from 5:15pm to 5:30pm indicates that the GARZ model has less prediction accuracy when the property value $w$ computed based on sensor data under a free-flow condition is later used in predicting density and velocity under congestion. This is due to the fact that the spread of data in the low density area is very small, and a small measurement noise might result in a property value $w$ computed by the GARZ that is very different from the true value. It would not cause very inaccurate prediction when using this inaccurately computed $w$ to predict free-flow traffic conditions, since in the low density regime the flow--density relationship varies little with property $w$. However, when the GARZ model uses the inaccurate $w$ to predict traffic conditions under congestion, a small error in $w$ would potentially result in rather large prediction error, given that the traffic condition varies more significantly with $w$ under congestion. Indeed, in the NGSIM data from 5:15pm to 5:30pm, the property values computed from the upstream side (relatively light traffic) are later propagated to the middle part of the road segment (more congested) and are used to predict traffic conditions in the middle area with more congestion. Consequently, the GARZ model has large error than the CGARZ model in this dataset.

\item The ARZ model has very poor prediction accuracy for the NGSIM dataset. This is due to the fact that some FD curves of the ARZ model are unrealistic in the congested regime, and the presence of significant congestion in the NGSIM dataset. Among the three studied time intervals, more congestion is observed during 5:00pm-5:15pm and 5:15pm-5:30pm. As a consequence, the prediction errors of the ARZ model during these two time intervals are significantly larger than for the other models.

\item Although the PT model outperforms the LWR and the ARZ models, especially when significant congestion is observed, it has larger prediction errors compared to the CGARZ model. Moreover, the numerical solver of the PT model \cite{blandin2013phase,BlandinWorkGoatinPiccoliBayen2011} is much more complicated than the 2CTM. 
\end{enumerate}

\subsection{Model Validation with the RTMC Sensor Data}\label{sec:validation_rtmc}
In order to better understand the distribution of the model prediction errors, we test the models on a larger dataset obtained from the RTMC sensor data \cite{TrafficMnDOT}. While the dataset covers a significantly longer time interval compared to NGSIM, it has the distinct disadvantage that the true traffic state is unknown. Consequently the model predictions are be compared to noisy measurement data not used in the model in a process described in detail next. 
\subsubsection{Analysis Setup}\label{sec:setup_rtmc}
In the model validation with the RTMC sensor data, a four-lane freeway segment of length 1.214km on the I-35W, Minneapolis is considered, which includes three sensors, with sensor~1 and sensor~3 located at the ends of the road segment, and sensor~2 located roughly in the middle of the study area. Following the test framework described in \cite{fan2014comparative,FanSeibold2013,BlandinBrettiCutoloPiccoli2009}, a three-detector test problem \cite{Daganzo1997} is solved. In this process, the accuracy of a model is quantified by the disparity between the measured states (traffic density and velocity) at sensor~2 and the model predictions at the same position, computed by numerically solving \eqref{eq:generic} with boundary conditions generated from the sensors~1 and~3. 

In this analysis, we use traffic data of 74 weekdays between 01/01/2003 and 04/14/2003. A small subset of those days is used for the model calibration. Specifically, the flow--density data from sensor~2 at days $d \in \{1,7,13,19,25,31,\allowbreak 37,43,49,55,61,67,73\}$ are used. The reasons for choosing those 13 days for calibration are: (\textit{i}) given that the 74 days are all weekdays, selecting one every other six days avoids any weekly patterns potentially embedded in the data, and (\textit{ii}) removing correlations between consecutive days due to similar weather, events, etc. Because the discrepancies of the various traffic models are of particular importance in the congested regime (e.g., the second-order models describe the spread of the flow--density data in the congested regime while the first-order model, LWR, does not), the model validation is conducted for the rush hour 4:00pm--5:00pm. After excluding the days used for calibration, we select 37 days among the remaining 61 days where the average density during 4:00pm--5:00pm is above 20 veh/km/lane for model validation (on the other days, not enough congestion was observed to obtain meaningful results). To generate a realistic initial state from the boundary data given at sensors~1 and~3, an initialization phase of five minutes (i.e., 4:00pm--4:05pm) is applied (see \cite{FanSeibold2013}).

\begin{figure}
\begin{center}
  \subfloat[]{\label{fig:fig_cgarz_eq_cali_minn_c}\includegraphics[width=0.4\textwidth]{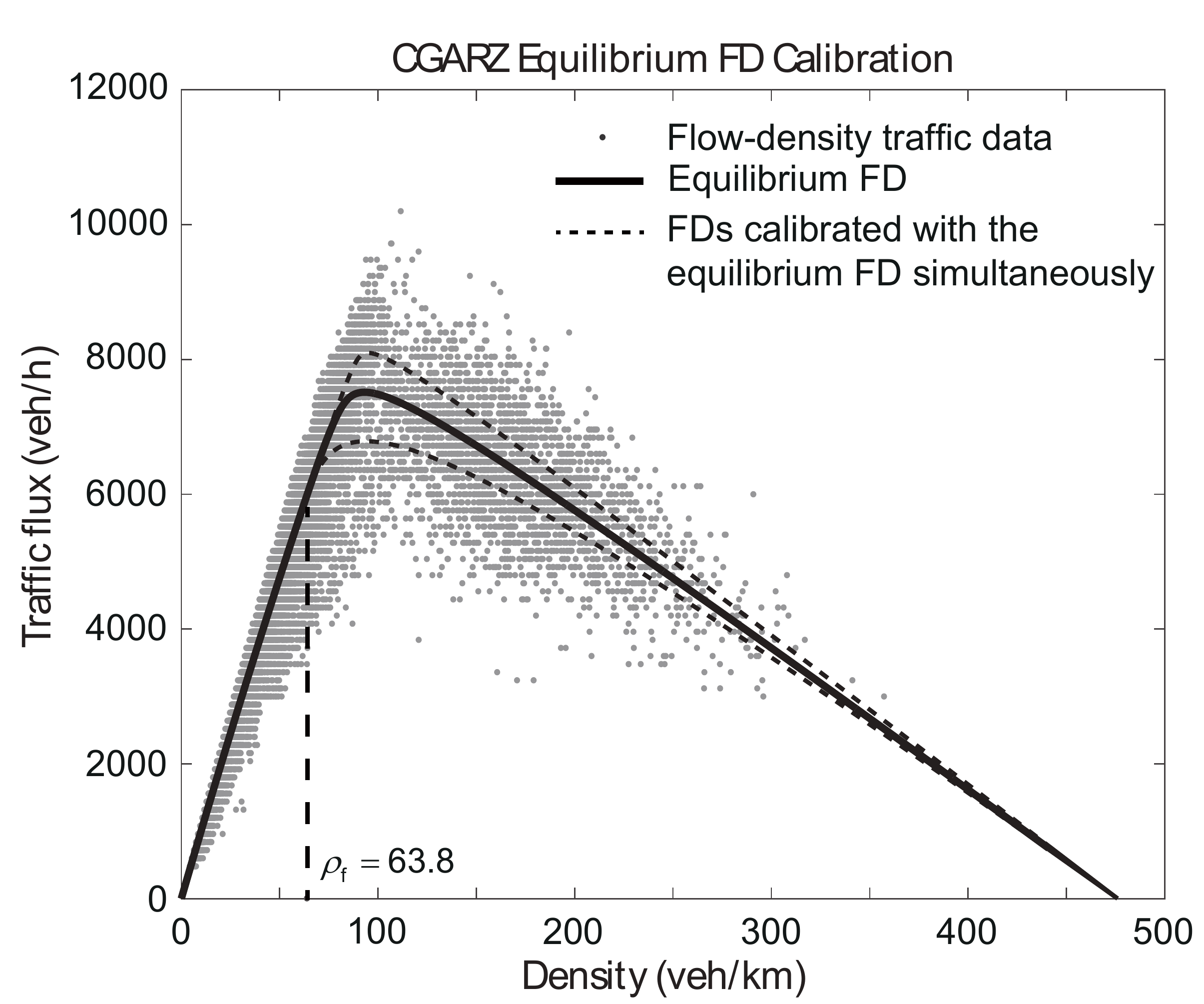}}
~\qquad
  \subfloat[]{\label{fig:fig_cgarz_cali_minn_c}\includegraphics[width=0.4\textwidth]{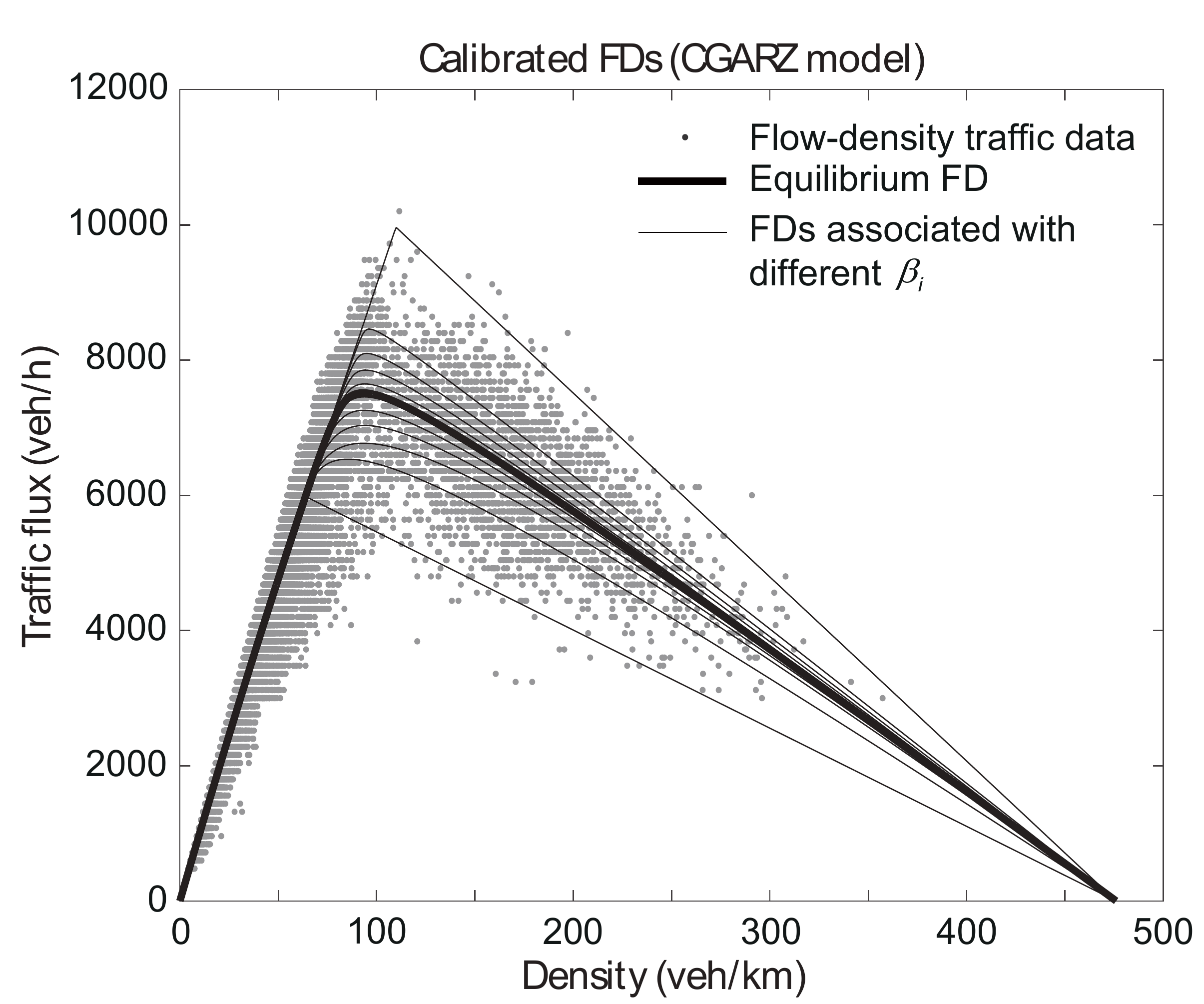}}\\
  \subfloat[]{\label{fig:fig_garz_cali_minn_c}\includegraphics[width=0.4\textwidth]{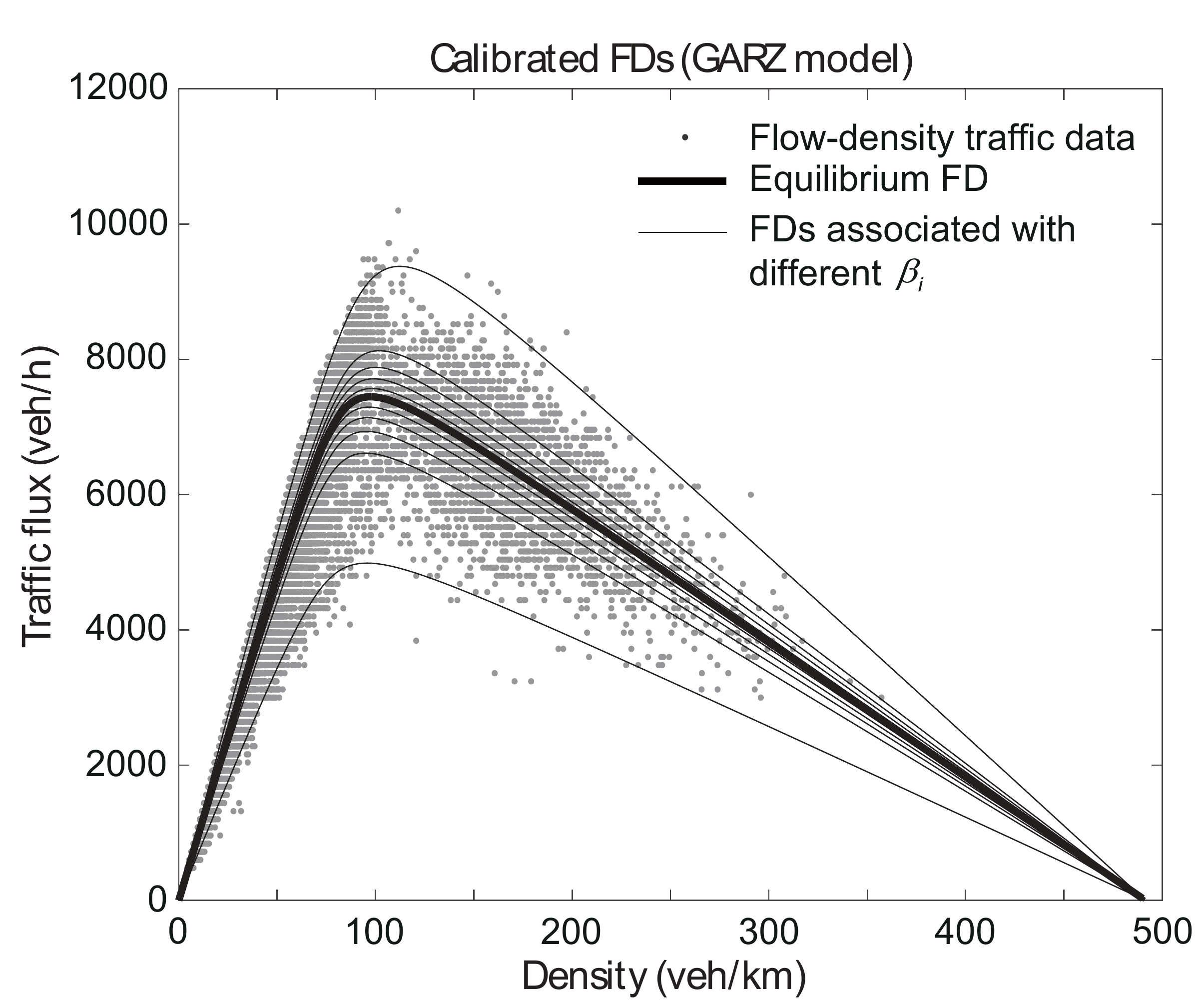}}
~\qquad
  \subfloat[]{\label{fig:fig_pt_cali_minn_c}\includegraphics[width=0.4\textwidth]{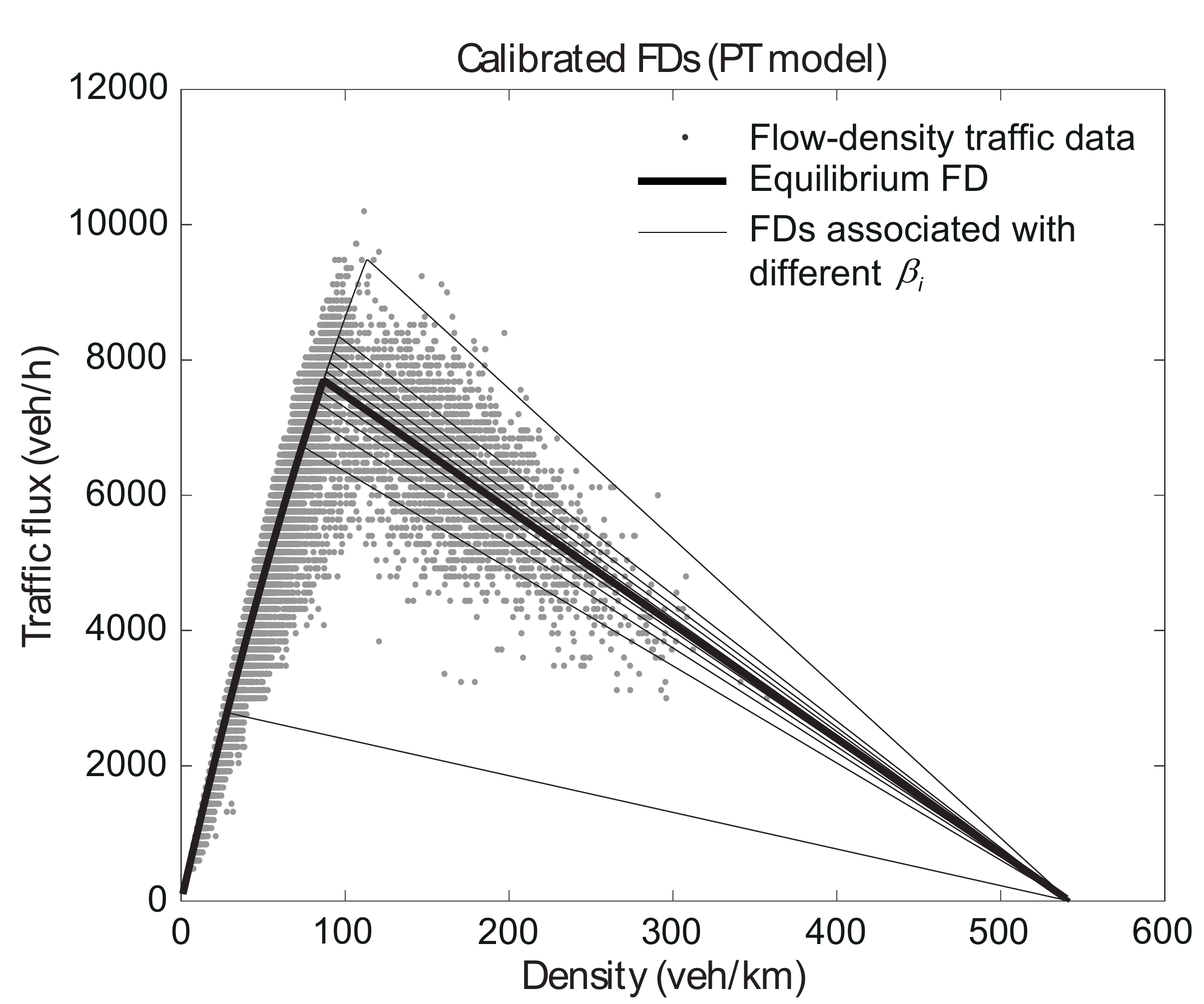}}
\caption{Model calibration using the flow--density data provided by the RTMC dataset.(a): The calibrated equilibrium FD of the CGARZ model and the two FDs (corresponding to $\beta_{1,\text{eq}}=0.2$ and $\beta_{2,\text{eq}}=0.8$) calibrated together with the equilibrium FD. (b-d): The calibrated family of fundamental diagrams of the CGARZ (b), the GARZ (c) and the PT (d) models. The calibrated family of FDs are plotted for a selection of $\beta$-values (i.e., $\beta_i$ in \eqref{eq:WLSQ_beta_family} with $i$ given in \eqref{eq:WLSQ_beta_family_plot}). }
\label{fig:fig_minn_cali_c}
\end{center}
\end{figure}

The calibrated parameters for all the traffic models are detailed in Table~\ref{tb:parameters_rtmc} in the Appendix. To obtain the free-flow threshold density $\rho_{\text{f}}$ for the CGARZ model in Step~1, we calibrate two curves together with the equilibrium curve (as shown in Figure~\ref{fig:fig_cgarz_eq_cali_minn_c}) using weight parameters $\beta_{1,\text{eq}}=0.2$, $\beta_{2,\text{eq}}=0.8$, and set $\tau=300$ veh/hr/lane. Compared to the loop detector data in Figure~\ref{fig:fig_cgarz_rtmc_cali}, the RTMC dataset has much more spread in the low density regime. Hence, two curves closer to each other are calibrated together with the equilibrium FD in order to avoid the calibrated $\rho_{\text{f}}$ being unreasonably small, which is different from the case in Figure~\ref{fig:fig_cgarz_eq_cali_ngsim_c} where $\beta_{1,\text{eq}}=0.001$, $\beta_{2,\text{eq}}=0.999$. In the sensitivity analysis of the shrinkage operator, the calibrated $\rho_{\text{f}}$ stays between 70 and 80 veh/km/lane for all $300\le \tau \le 700$ veh/hr/lane, which again shows that the calibration result is rather insensitive to the $\tau$ values in the shrinkage operator. In Step~2, a family of 100 curves is calibrated for the CGARZ (Figure~\ref{fig:fig_cgarz_cali_minn_c}), GARZ (Figure~\ref{fig:fig_garz_cali_minn_c}) and the PT models (Figure~\ref{fig:fig_pt_cali_minn_c}), where the family of $\beta_i$ is given in \eqref{eq:WLSQ_beta_family}. In Step~3, a polynomial regression is applied to construct the relationship between $w$ and the obtained parameters.


\subsubsection{Error Metric}\label{sec:error_metric_rtmc}
For the RTMC data, we compare the density and velocity measured at sensor~2 with the corresponding values predicted by the traffic models. Let $x_{s2}$ be the position of sensor~2, and $n_{\text{lane}}$ be the number of lanes. Then the density and velocity errors at location $x_{s2}$ and time $t\in[0,T]$ are given by:
\begin{equation*}
\label{eq:error_measure_s2}
\begin{split}
E_{\rho}(x_{s2},t) &= \frac{\left|\rho^\text{model}(x_{s2},t)-\rho^\text{sensor}(x_{s2},t)\right|}{n_{\text{lane}}},\\
E_{v}(x_{s2},t) &= \left|v^\text{model}(x_{s2},t)-v^\text{sensor}(x_{s2},t)\right| ,
\end{split}
\end{equation*}
where $\rho^\text{model}(x_{s2},t)$ and $v^\text{model}(x_{s2},t)$ (resp.~$\rho^\text{sensor}(x_{s2},t)$ and $v^\text{sensor}(x_{s2},t)$) are the predicted (resp.\ measured) density and velocity at location $x_{s2}$ and time $t$. Hence, the temporal average error on day $d$ is computed as follows:
\begin{equation*}
\label{eq:error_rtmc_avg_t}
E^d_{\rho} = \frac{1}{T}\int_0^T E_{\rho}(x_{s2},t)\ud{t}\;,\quad
E^d_v = \frac{1}{T}\int_0^T E_{v}(x_{s2},t)\ud{t}\;.
\end{equation*}
Denote as $\mathcal{D}_{\text{val}}$ the set of days selected for validation, and $|\mathcal{D}_{\text{val}}|$ the total number of days in $\mathcal{D}_{\text{val}}$. As discussed above, we have $|\mathcal{D}_{\text{val}}|=37$. The average density and velocity errors with respect to all the selected days are:
\begin{equation*}
\label{eq:error_rtmc_avg_days}
E _{\rho}= \frac{1}{|\mathcal{D}_{\text{val}}|}\sum_{d\in \mathcal{D}_{\text{val}}} E^d_{\rho}\;,\quad
E_{v} = \frac{1}{|\mathcal{D}_{\text{val}}|}\sum_{d\in\mathcal{D}_{\text{val}}} E^d_{v}\;.
\end{equation*}

\subsubsection{Model Validation Results}\label{sec:results_rtmc}

\begin{figure}
\begin{center}
\includegraphics[width=\textwidth]{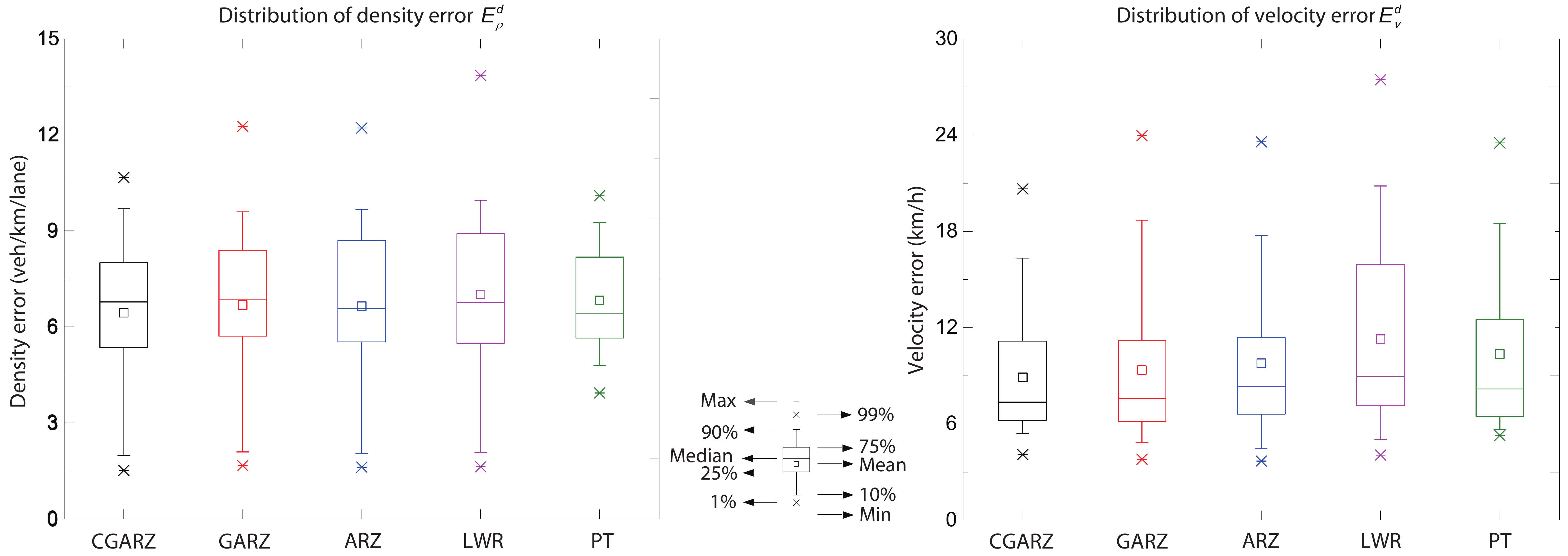}
\caption{Distribution of the density prediction error $E_{\rho}^d$ (a) and the velocity prediction error $E_v^d$ (b) of the considered models for the days $d\in\mathcal{D}_{\text{val}}$, where $\mathcal{D}_{\text{val}}$ is the set of days selected for model validation.}
\label{fig:error_rtmc}
\end{center}
\end{figure}

Figure~\ref{fig:error_rtmc} illustrates the distribution of the density error $E_{\rho}^d$ and the velocity error $E_{v}^d$ of all the considered models for $d\in \mathcal{D}_{\text{val}}$. The key findings regarding the prediction accuracies of the considered models are given as follows:
\begin{enumerate}[1.]
\item When the traffic is in free-flow, the model prediction accuracies for all considered models are close to each other and of good quality. Hence, the medians and the 25th percentile of all the models are similar, and the days with average errors $E_{\rho}^d$ and $E_{v}^d$ smaller than the median typically have a large portion of time between 4:00pm--5:00pm in free-flow conditions. 

\item Due to the apparent modeling difficulty to accurately predict the large variations of the flow--density relationship in the high density regime, the prediction errors of all the models increase as a larger amount of congestion is observed during 4:00pm--5:00pm. Because the second-order models, as well as the PT model, can capture the spread of flow--density data in congestion, it agrees with our expectation that the prediction errors of the CGARZ, GARZ, ARZ, and PT models are smaller than with LWR when there is more congestion during the studied time period. This fact is apparent in the larger 75th percentile and outliers of the LWR model compared to other models. As a consequence, the LWR model has the largest average errors $E_{\rho}$ and $E_{v}$.

\item The statistics of $E^d_{\rho}$ and $E^d_{v}$ show that the CGARZ model has the best prediction accuracy, in addition to its fundamental advantage that the model dynamics of the CGARZ model are substantially simpler in the free-flow regime compared to the GARZ and ARZ models.

\item The ARZ model also captures the spread of flow--density data. However, unlike the GARZ model, different maximum densities are obtained for different FD curves, and some of those densities are clearly unrealistic, as shown in \cite[Figure 3.4]{Fan2013}. Due to this modeling shortcoming of ARZ, it is not a surprise that the CGARZ model outperforms the ARZ model. Because the RTMC dataset exhibits not too much congestion (see the data points in Figure~\ref{fig:fig_minn_cali_c}), the CGARZ model only slightly outperforms the ARZ model. In contrast, in the NGSIM dataset, more congestion is present, and consequently the advantage of the CGARZ model over the ARZ model is quite apparent.

\item Compared to the GARZ and the ARZ models, the CGARZ model has the smallest 75th percentiles and the smallest maximum values of $E^d_{\rho}$ and $E^d_{v}$. Although the GARZ and ARZ models are able to describe the spread of traffic data in the free-flow regime, the property $w$, computed by the flow--density data pairs in the free-flow regime, is more likely to be inaccurate due to the less spread of data in the low density area and sensor measurement errors (see bullet point 2 in \ref{sec:results_ngsim}). As a consequence, compared to the CGARZ model which describes the flow--density relationship by a single curve in the low density regime, the GARZ and ARZ models can potentially provide very large prediction errors. In fact, on the days when the GARZ model produces high prediction errors, a large portion of the boundary flow--density data points are very close to the highest calibrated FD in the free-flow regime, which leads to very large property values $w$ computed by the inverse function $W(\cdot)$ introduced in Section~\ref{section:inverse}. However, it is likely that the true properties are not as large as the computed $w$, and the data points are close to the upper FD curve just due to the sensor error. As stated in bullet point 2 in \ref{sec:results_ngsim}, the inaccurately computed properties would result in large prediction errors especially when propagating to the congested areas.
\end{enumerate}

\section{Conclusion}
We have compared the CGARZ model with various traffic models that fit into the GSOM framework, as well as with the PT model. By collapsing the family of FD curves in the free-flow regime, the CGARZ model has simpler dynamics in the low density regime, compared to the GARZ and ARZ models. In addition, since the CGARZ model also belongs to the GSOM class, it avoids several complicated analysis and specific implementation steps required by the PT model. In order to numerically solve the models GSOM class, we have presented a 2CTM with clear physical interpretation, designed by analyzing the inflow and outflow of both traffic density $\rho$ and total property $y=\rho w$ on each cell of the road segment. To improve the predictive accuracy of the traffic models, a systematic approach to generate model parameters from historic data has been introduced, which is applicable to all the traffic models considered in this work. Model validations and comparisons are conducted on the NGSIM and the RTMC datasets, and the key findings are summarized as follows:
\begin{enumerate}[1.]
 \item While providing much simpler model dynamics in the free-flow regime (than other GSOMs), the CGARZ model maintains high prediction quality.
 
 \item Due to the unrealistic FD curves in the congested regime, the ARZ model can only provide competitive prediction quality when the traffic conditions are largely in free-flow.

 \item  Computing the property $w$ from flow--density data in the free-flow regime can potentially lead to large prediction errors (especially when predicting traffic conditions under congested status), because the data spread in the low density area is small and the computed $w$ is very sensitive to noise. Hence, the GARZ and the ARZ models occasionally produces very large prediction errors. The CGARZ model remedies this problem.

 \item Due to the capability of the second-order models to describe the variations of the flow--density relationship in the congested regime, second-order models in general have better prediction accuracy compared to the first-order LWR model.

\end{enumerate}

The present study demonstrates that the CGARZ model combines the advantages of accuracy in the congested regime and simplicity in the free-flow regime. It is therefore a good candidate for traffic state estimation, prediction, and control of contemporary traffic flow. The modeling framework presented here can also be extended to triangular fundamental diagrams.

Looking forward, the proposed modeling framework of the CGARZ model will become even more important with the introduction of smart (e.g., connected or automated) vehicles on the roadways. In this setting, the property $w$ can be used to represent the fraction of flow represented by smart vehicles ($w=0$ means all vehicles are human-operated, and $w=1$ means all vehicles are smart). In free-flow, all vehicles drive at the same speed as dictated by the speed limit, hence a single collapsed FD curve arises. In turn, in congested flow, the ability of the smart vehicles to follow other vehicles more closely (shorter reaction times and non-local information) increases the density associated to a given average speed. Hence a family of FD curves is obtained in congestion, parameterized by $w$. In other words, the introduction of smart vehicles will shift real traffic dynamics naturally even closer to a collapsed GARZ framework.

\newpage
\section*{Appendix}
\renewcommand{\thetable}{A.{\arabic{table}}}

\begin{table}[h]
  \centering
  \begin{threeparttable}[b]
  \begin{tabular}{c|c|c|c}
  \toprule
  & \multicolumn{1}{c}{Step~1} & \multicolumn{1}{c}{Step~2} & \multicolumn{1}{c}{Step~3}\\ \hline
  \multirow{4}{*}{} & \multicolumn{3}{c}{Parameters used in the calibration procedure}\\\cline{2-4}
  &$m_{\text{eq}}=2$, $\tau=400$,& $\beta_i=10^{-3}+\frac{0.998}{99}(i-1)$&\multirow{3}{*}{$k=5$}\\
  &$\beta_{1,\text{eq}}=0.001$, $\beta_{2,\text{eq}}=0.999$& for $i \in \{1,\cdots,100\}$ & \\
  &(only for the CGARZ)& &\\\hline
  & \multicolumn{3}{c}{Calibrated parameters}\\\cline{2-4}
  \multirow{3}{*}{CGARZ} & $v_{\max}=73.5$, $\rho_{\text{f}}=75.9$,& $\sigma_{\beta_i} \in [5.48, 22.37]$,& $a_{\sigma}=(30.3,-0.0014)$,\\[-1ex]
  & $\tilde{\rho}_{\max}=1399.9$, $\rho_{\max}=801.5$,&$\mu_{\beta_i} \in [69.4, 179.0]$,& $a_{\mu}=(-51.8, 0.02)$\\[-1ex]
  &$\sigma_{\text{eq}}=15.9$, $\mu_{\text{eq}}=129.8$& (see Figure~\ref{fig:fig_cgarz_cali_ngsim_c}) & \\
  \multirow{4}{*}{GARZ}& &$\alpha_{\beta_i} \in [1116.4, 2963]$& $a_{\alpha}=10^3\cdot(-1.9,0.05),$\\[-1ex]
  &\multirow{1}{*}{$\rho_{\max}=809.3$, $\alpha_{\text{eq}}=1450.9$}&$\lambda_{\beta_i} \in [13.95, 26.46]$ & $a_{\lambda}=(43.25,-0.28),$ \\[-1ex]
  &\multirow{1}{*}{$\lambda_{\text{eq}}=24.1$, $p_{\text{eq}}=0.16$}&$p_{\beta_i} \in [0.13, 0.21]$&$a_p=(-0.2, 0.005)$\\[-1ex]
  &&(see Figure~\ref{fig:fig_garz_cali_ngsim_c})& \\
  ARZ & Same as the GARZ & None & None\\
  LWR & Same as the GARZ & None & None\\
  \multirow{2}{*}{PT} & $v_{\max}=74.3$, $\tilde{\rho}_{\max}=884.0$&$w_{\beta_i}\in[5596.2,11759]$ &\multirow{2}{*}{None}\\[-1ex]
  &$\rho_{\max}=884.0$, $w_{\text{eq}}=8923.5$ &(see Figure~\ref{fig:fig_pt_cali_ngsim_c})&\\
  \bottomrule
  \end{tabular}
  \end{threeparttable}
  \vskip.4cm
\caption{Parameters of all the considered models calibrated using the PeMS loop detector data near the freeway segment generating the NGSIM data. The calibrated parameters are used in the model validations with the NGSIM sensor data, where densities are given in veh/km/lane, and velocities are given in km/hr.}
 \label{tb:parameters_pems}
\end{table}

\begin{table}[h]
  \centering
  \begin{threeparttable}[b]
  \begin{tabular}{c|c|c|c}
  \toprule
  & \multicolumn{1}{c}{Step~1} & \multicolumn{1}{c}{Step~2} & \multicolumn{1}{c}{Step~3}\\ \hline
  \multirow{4}{*}{} & \multicolumn{3}{c}{Parameters used in the calibration procedure}\\\cline{2-4}
  &$m_{\text{eq}}=2$, $\tau=300$,& $\beta_i=10^{-3}+\frac{0.998}{99}(i-1)$&\multirow{3}{*}{$k=5$}\\
  &$\beta_{1,\text{eq}}=0.2$, $\beta_{2,\text{eq}}=0.8$& for $i \in \{1,\cdots,100\}$ & \\
  &(only for the CGARZ)& &\\\hline
  & \multicolumn{3}{c}{Calibrated parameters}\\\cline{2-4}
  \multirow{3}{*}{CGARZ} & $v_{\max}=100.8$, $\rho_{\text{f}}=63.8$,& $\sigma_{\beta_i} \in [0.001, 10.987]$,& $a_{\sigma}=(26.0,-0.003)$,\\[-1ex]
  & $\tilde{\rho}_{\max}=901.6$, $\rho_{\max}=476.1$,&$\mu_{\beta_i} \in [63.8, 109.8]$,& $a_{\mu}=(-29.3, 0.015)$\\[-1ex]
  &$\sigma_{\text{eq}}=6.4$, $\mu_{\text{eq}}=83.1$& (see Figure~\ref{fig:fig_cgarz_cali_minn_c})& \\
  \multirow{4}{*}{GARZ}& &$\alpha_{\beta_i} \in [896.9, 1701.1]$& $a_{\alpha}=10^3\cdot(8.60, 0.18),$\\[-1ex]
  &\multirow{1}{*}{$\rho_{\max}=491.5$, $\alpha_{\text{eq}}=1033.6$}&$\lambda_{\beta_i} \in [20.13, 36.20]$ & $a_{\lambda}=(-156.6, 4.05),$ \\[-1ex]
  &\multirow{1}{*}{$\lambda_{\text{eq}}=28.3$, $p_{\text{eq}}=0.17$}&$p_{\beta_i} \in [0.19, 0.45]$&$a_p=(0.31, -0.004)$\\[-1ex]
  &&(see Figure~\ref{fig:fig_garz_cali_minn_c}) & \\
  ARZ & Same as the GARZ & None & None\\
  LWR & Same as the GARZ & None & None\\
  \multirow{2}{*}{PT} & $v_{\max}=104.9$, $\tilde{\rho}_{\max}=562.8$&$w_{\beta_i}\in[2785.8,9494.0]$ &\multirow{2}{*}{None}\\[-1ex]
  &$\rho_{\max}=542.1$, $w_{\text{eq}}=7701.7$ &(see Figure~\ref{fig:fig_pt_cali_minn_c})&\\
  \bottomrule
  \end{tabular}
  \end{threeparttable}
  \vskip.4cm
\caption{Parameters of all the considered models calibrated using the RTMC data on the selected days $d\in\{(d-1) \mod 6=0|1\leq d \leq 74\}$, where densities are given in veh/km/lane, and velocities are given in km/hr.}
 \label{tb:parameters_rtmc}
\end{table}

\newpage

\section*{Bibliography}
\bibliographystyle{IEEEtran}
\bibliography{references_complete}

\end{document}